\documentclass[10pt,journal,  oneside, twocolumn]{IEEEtran}

\usepackage{multirow}
\usepackage{latexsym}
\usepackage{graphicx}
\usepackage{float}
\usepackage{amsmath}
\usepackage{amsthm}
\usepackage{lipsum}
\usepackage{subfig}
\usepackage{graphicx}
\usepackage{authblk}
\usepackage{bm}
\usepackage{booktabs}
\usepackage{amsthm}
\usepackage[section]{placeins}
\usepackage{soul}

\usepackage{colortbl}

\makeatletter  
\newif\if@restonecol  
\makeatother

\usepackage[linesnumbered,ruled,vlined]{algorithm2e}
\usepackage{algpseudocode}  
\usepackage{amsmath}

\usepackage{amssymb}
\DeclareMathOperator*{\argmax}{argmax}

\usepackage{mathrsfs}
\usepackage{subfig}
\usepackage{caption}
\newcounter{problem}

\SetKwInput{KwInitialize}{Initialize}

\begin{document}

\title{Digital Twin-based SIM Communication and Flight Control for Advanced Air Mobility}


\author{Kai Xiong,~\IEEEmembership{Member,~IEEE}, Zhen Chen, Juefei Xie, Supeng Leng,~\IEEEmembership{Member,~IEEE}, Chau Yuen,~\IEEEmembership{Fellow,~IEEE}

\thanks{

K. Xiong, Z. Chen, J. Xie, and S. Leng are with School of Information and Communication Engineering, University of Electronic Science and Technology of China, Chengdu, 611731, China; and, Shenzhen Institute for Advanced Study, University of Electronic Science and Technology of China, Shenzhen, 518110, China.
}


\thanks{
C. Yuen is with School of Electrical and Electronics Engineering, Nanyang Technological University, 639798, Singapore. 
}






\thanks{The corresponding author is Supeng Leng, email: \{spleng, xiongkai\}@uestc.edu.cn}
}


\maketitle



\begin{abstract}
Electric Vertical Take-off and Landing vehicles (eVTOLs) are driving Advanced Air Mobility (AAM) toward transforming urban transportation by extending travel from congested ground networks to low-altitude airspace.
This transition promises to reduce traffic congestion and significantly shorten commute times. 
To ensure aviation safety, eVTOLs must fly within prescribed flight corridors. These corridors are managed by ground-based Air Traffic Control (ATCo) stations, which oversee air-ground communication and flight scheduling. However, one critical challenge remains: the lack of high rate air-ground communication and safe flight planning within these corridors.
The introduction of 6G-oriented Stacked Intelligent Metasurface (SIM) technology presents a high rate communication solution. With advanced phase-shifting capabilities, SIM enables precise wireless signal control and supports beam-tracking communication with eVTOLs. Leveraging this technology, we propose a Composite Potential Field (CPF) approach. This method dynamically integrates target, separation, and communication fields to optimize both SIM communication efficiency and flight safety.
Simulation results validate the effectiveness of this DT-based approach. Compared to the potential field flight control benchmark, it improves the transmission rate by 8.3\%. Additionally, it reduces flight distance deviation from the prescribed corridor by 10\% compared to predetermined optimization methods.

\end{abstract}



\IEEEpeerreviewmaketitle

\section{Introduction}

\IEEEPARstart{A}{dvanced} Air Mobility (AAM) is rapidly transitioning from concept to reality \cite{2023Advanced}. Spearheaded by initiatives such as NASA’s AAM framework, this vision re-imagines transportation by utilizing low-altitude airspace to move people and goods efficiently. By 2030, the AAM industry is projected to flourish, driven by advancements in Communication, Computing, and Control (3C) technologies \cite{ICAAMap2021}.

At the heart of AAM lies Air Traffic Control (ATC), scheduled by ground-based ATCo stations \cite{9120233Bapinar}. These stations are responsible for generating optimal ATC plans and broadcasting them to eVTOLs through air-ground communication. 
Moreover, the Federal Aviation Administration's (FAA) Innovate28 (I28) initiative outlines operational models for eVTOLs, emphasizing safety through prescribed, non-overlapping air corridors \cite{FAAd93329}. These corridors form the backbone of low-altitude airspace, ensuring eVTOL safe navigation. However, two critical challenges remain, maintaining efficient air-ground communication and generating optimal flight control plans along the prescribed air corridor.

Existing mobile networks, primarily designed for ground users, face limitations in serving low-altitude aircraft like eVTOLs. Antenna inclination constraints often place eVTOLs outside the effective coverage zones.
While deploying aerial base stations could enhance air-ground communication, this solution is costly and may cause significant air-ground interference. Alternatively, retrofitting ground base stations to improve air-ground signal strength offers a more feasible option.
Fortunately, emerging technologies, such as 6G-oriented Holographic Multiple Input Multiple Output (HMIMO), promise to address these limitations by enabling customizable wireless propagation toward low-altitude airspace \cite{9136592}. By integrating ground and aerial networks through low-cost metasurfaces, HMIMO significantly reduces deployment costs. However, single-layer metasurfaces currently limit HMIMO systems. Stacked Intelligent Metasurfaces (SIM), which cascade multiple metasurfaces, offer enhanced flexibility in beamforming, making them easily signal track for dynamic eVTOLs \cite{lin10445200}.

Regarding optimal flight control, urban airspace presents unique challenges for eVTOLs, including movable obstacles such as aircraft and birds. Flight control must dynamically adapt to these challenges while ensuring safe and efficient aviation. Typically, flight control is categorized into two methods, strategic scheduling, performed before takeoff, and tactical scheduling, performed during flight \cite{Robert2355}.
Strategic scheduling ensures conflict-free navigation by establishing safe prescribed corridors before takeoff. Tactical scheduling, on the other hand, adjusts flight parameters in real time to respond to environmental changes, such as unexpected obstacles or wind patterns. This paper focuses on tactical scheduling within prescribed flight corridors, complementing pre-established strategic plans.

The Artificial Potential Field (APF) method is a common approach to tactical scheduling. APF creates attractive fields to guide eVTOLs toward their targets and repulsive fields to avoid obstacles \cite{7737540McIntyre}. By combining multiple potential fields—for collision avoidance, formation flying, and trajectory optimization—eVTOLs can safely navigate in real time \cite{Selvam9448937}. However, existing APF methods fail to achieve both safe navigation and communication efficiency within prescribed corridors. Moreover, the frequent APF parameter adjustments required for smooth flight through these corridors complicate real-time flight optimization. It also needs high-speed and stable air-ground communication support to frequently update the changed APF parameters.


However, both SIM-based beam tracking and eVTOL tactical scheduling demand substantial computational resources for real-time optimization. 
Onboard systems often lack the capacity to meet these demands. Digital Twins (DTs) provide a powerful alternative \cite{2017Digital}. DTs replicate physical systems in a virtual environment, allowing for efficient testing, analysis, and optimization without risking real-world operations. In our proposed framework, DTs are used to optimize SIM beamforming and eVTOL flight control through iterative DT deductions. This reduces costs, risks, and computational demands while enabling precise adjustments.
By offloading computational tasks to ground-based ATCo stations, the DT framework minimizes onboard processing requirements. DT synchronization between the physical eVTOLs and the digital presentations ensures the accuracy of flight parameters and communication settings.
But, frequent DT synchronization still generates significant communication overhead.
It is necessarily to design appropriate parameters for DT synchronization.

This paper introduces a DT-based framework for jointly optimizing SIM communication and eVTOL flight control. 
The proposed framework addresses two core challenges, ensuring robust air-ground communication and achieving efficient flight within prescribed corridors. By offloading the optimization tasks to the DT system, the framework minimizes the processing burden of eVTOLs, reducing operational risks while improving communication and aviation efficiency. The optimized parameters for communication and flight are pre-calculated in the DT environment, and only key hyperparameters are transmitted during operation, significantly reducing the DT synchronization overheads. The primary contributions of this study are as follows:

\begin{itemize}


\item We propose a novel two-step iterative method to optimize SIM-based air-ground communication between ATCo stations and eVTOLs. 
This method dynamically beam-tracks the trajectory of eVTOLs within prescribed corridors, maximizing the air-ground transmission rate over time while accounting for diverse SIM configurations.
Simulation results demonstrate that SIM-based communication significantly outperforms traditional Multi-Input Multi-Output (MIMO) systems in terms of transmission rate.
The improvement is attributed to the configurable multi-layer phase shift component, which enhances air-ground beamforming for dynamic eVTOLs.



\item We develop a Deep Q-learning Network (DQN)-based composite potential field method that maintains safe aviation and communication connectivity. 
By dynamically adjusting hyperparameters for target tracking, safe separation, and communication connectivity fields, the CPF method allows eVTOLs to navigate complex corridors safely and efficiently. 
This flight control mechanism accommodates real-time changes in environmental conditions.
Additionally, the CPF scheme minimizes the data required for DT synchronization by transmitting only key hyperparameters of the potential fields. It is suitable for real-world deployment in dynamic urban airspaces.


A DT-based optimization framework is proposed to jointly manage SIM communication and eVTOL flight control. The system incorporates two DTs: (i) the SIM DT, which represents the physical SIM antenna at the ATCo station, and (ii) the eVTOL DT, which mirrors the dynamic behavior and flight characteristics of eVTOLs.
By offloading complex computational tasks to these virtual representations, the framework decouples flight control optimization from onboard systems. This separation enables eVTOLs to autonomously maintain connectivity, navigate safely, and avoid collisions based on pre-calculated trajectories. The DT-based approach leverages the global situational awareness and computational power of ATCo servers, improving operational safety and efficiency in low-altitude airspaces while reducing onboard processing demands.

\item We propose a DT-based optimization framework that integrates SIM communication and eVTOL flight control. The system incorporates two DTs, the SIM DT, representing the physical SIM antenna, and the eVTOL DT, mirroring the dynamic behavior of eVTOLs.
By offloading optimization tasks to these virtual representations, these DTs
enable eVTOLs to autonomously maintain connectivity, navigate safely, and avoid collisions based on pre-calculated trajectories. The DT-based approach leverages the global situational awareness and computational power of ATCo servers, improving operational safety and efficiency in airspaces while reducing onboard processing demands.

\end{itemize}
The rest of the paper is organized as follows:
Section II reviews related work.
Section III presents the system model.
Section IV details our proposed optimization solutions for SIM communication and eVTOL flight control.
Section V evaluates performance through simulations and analysis.
Section VI concludes the paper.

\section{Related Work}

Engineers and researchers are actively developing AAM to create efficient and reliable aerial transportation systems. Technologies such as Stacked Intelligent Metasurfaces (SIM) and Digital Twins (DT) are increasingly recognized for their potential to enhance air-ground communication and improve eVTOL aviation safety. This section provides an overview of existing research on flight control, SIM communication, and DT technology.

\begin{figure*}[h]
\centering
     \includegraphics[width=.8\textwidth]{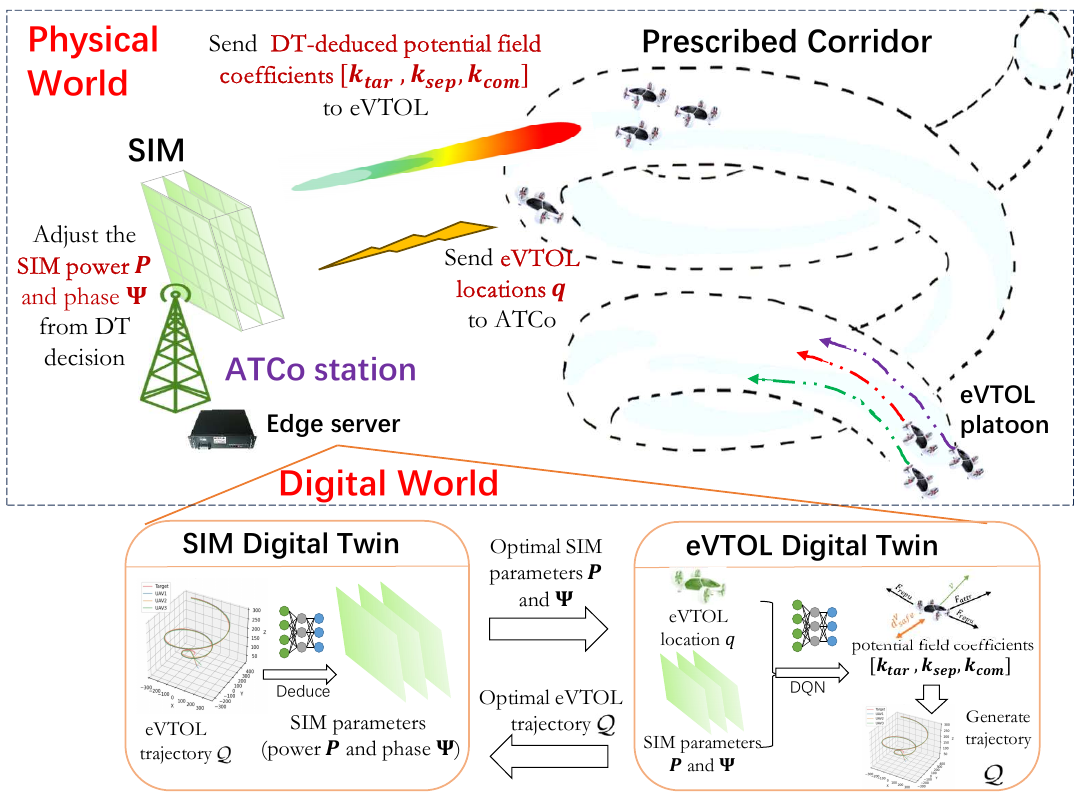} 
     \caption{DT-based SIM communication and eVTOL flight Optimization.} 
\label{system_model}
\end{figure*}

\subsection{Communication and Flight Control Optimization}
Connected aircraft, including Unmanned Aerial Vehicles (UAVs) and eVTOLs, are emerging as key components of 6G communication networks due to their ability to provide seamless air-ground service \cite{8758975}. However, ensuring reliable communication requires precise flight control and antenna configurations. For instance, when an eVTOL moves out of the current antenna beam’s coverage, the ATCo station must dynamically generate a new beamforming to track its updated position.

Urban low-altitude airspaces present additional challenges, including movable obstacles like drones/birds that create risks of collision and signal blocking. Wu \textit{et al.} \cite{8247211} addressed these issues by optimizing eVTOL flight planning and user association to maximize throughput. Zeng \textit{et al.} \cite{7888557} proposed circular eVTOL trajectories to improve energy efficiency by accounting for both propulsion energy consumption and throughput. However, these studies largely ignore the constraints of prescribed air corridors mandated by aviation authorities. This paper focuses on maximizing transmission throughput while ensuring safe eVTOL aviation along prescribed flight corridors.

\subsection{Stacked Intelligent Metasurface}
Reconfigurable Intelligent Surfaces (RIS) are gaining attention for their ability to improve communication quality of service (QoS) using cost-effective metasurfaces \cite{9410457Chongwen}. When metasurfaces function as transceivers—embedding energy-intensive radio frequency circuits and signal processing units—they are termed "active RIS." Active RIS is a natural evolution of conventional massive MIMO systems, incorporating increasingly dense, software-controlled antenna elements onto two-dimensional metasurfaces. This approach is also referred to as Holographic MIMO (HMIMO).

HMIMO systems dynamically control signal propagation through phase shift adjustments, enabling efficient and cost-effective beamforming. However, the use of single-layer metasurfaces in current HMIMO systems limits their flexibility for beamforming and interference suppression \cite{8930608}. SIM technology addresses this limitation by employing multi-layer metasurfaces, which provide greater degrees of freedom for creating complex beamforming patterns \cite{10279173}. While SIM is a promising technology for 6G communication, its application to low-altitude air-ground communication for flight control remains underexplored.


\subsection{Digital Twin}
Digital Twin (DT) technology has garnered significant attention across various industries, including manufacturing, healthcare, and aviation. Barricelli \textit{et al.} \cite{8901113} examined DT’s role in these fields, emphasizing its potential to enhance system efficiency and safety. In the automotive sector, Biesinger \textit{et al.} \cite{8932144} explored how DT can streamline integration and planning processes. Similarly, in UAV systems, DT has been applied to optimize flight control and navigation for both single UAVs \cite{Bae2022AutomaticFP} and UAV swarms \cite{10045049}.
Moreover, recent studies have demonstrated how integrating DT with reinforcement learning can reduce resource consumption in UAV systems \cite{Kai10117556}. Building on this foundation, this paper investigates the potential of combining communication optimization and flight control within a DT-based eVTOL system.

Despite significant progress in UAV/eVTOL communication and flight control research, a critical gap remains: the joint optimization of communication and flight trajectory under the constraints of prescribed flight corridors. Most existing studies focus on either communication or flight control independently, often overlooking the interconnected challenges posed by air corridor restrictions.
This paper addresses this gap by proposing a DT-based framework that integrates SIM communication and composite potential field flight control. The proposed system offers a unified approach to ensure both communication efficiency and safe aviation along prescribed air corridors.

\section{System Architecture}

The primary aim of this paper is to design a novel DT-based system that provides advanced communication and flight control technologies for efficient eVTOL navigation in a prescribed corridor. 
As illustrated in Fig.~\ref{system_model}, the proposed DT-based AAM system consists of two types of DTs: (i) SIM DT and (ii) eVTOL DT. The SIM DT is responsible for optimizing the phase shift and transmission power of multiple metasurfaces based on the flight trajectory generated by the eVTOL DT. In turn, the eVTOL DT refines flight control parameters using the updated phase shift and transmission power provided by the SIM DT.

When an eVTOL requests access to an air corridor, the ATCo station receives a message containing the eVTOL's current dynamic information (position $q$, velocity $v$). Using this data, the ATCo station creates a corresponding eVTOL DT on its edge server. The edge server also has direct access to the ATCo station's SIM antenna configurations, including phase shifts and transmission power levels, enabling the server to construct a SIM DT seamlessly.

The SIM DT ($DT^S$) and eVTOL DT ($DT^e$) operate iteratively, each leveraging the updated outputs of the other to optimize SIM parameters (phase shift and transmission power) and eVTOL trajectories. Once the iterative optimization process is complete, the DT-derived communication and flight control parameters are transmitted to the physical entities. Specifically, the SIM antenna configurations are updated via the ATCo station's wired circuits, while the optimized flight control parameters are communicated to the eVTOLs through SIM-based air-ground communication. These updated parameters ensure efficient navigation for the short-term future.
To guide the physical eVTOL throughout its corridor aviation, the ATCo station must continuously synchronize the DT-derived parameters with the physical entities. This continuous DT synchronization places high demands on air-ground communication systems, requiring robust and reliable performance.

The proposed communication scheme is to utilize SIM technology to enhance air-ground communication between the ATCo station and multiple eVTOLs. The SIM antenna consists of $L$ metasurface layers, each containing $K$ meta-atoms.
These meta-atoms can impose configurable phase shifts on the electromagnetic waves they transmit \cite{2020Beamforming}. By adjusting the phase shifts of individual meta-atoms, the SIM performs precise downlink beamforming for $M$ eVTOLs \cite{2022A}.
Here, key configurable parameters of the SIM include:
The phase shift matrix $\bm{\Psi}$, which determines the phase shifts of the $L\times K$ meta-atoms.
The transmission power allocation $\bm{P}$, which optimizes power distribution for the $M$ eVTOLs.
These configurable parameters allow the SIM to dynamically adapt to changing communication environments, ensuring efficient air-ground communication.

\subsection{Digital Twin Modeling}
The Digital Twin (DT) modeling framework in this paper builds upon our previous work \cite{Kai10117556}. The eVTOL Digital Twin ($DT^e$) digitally replicates the key features of the physical eVTOL, representing its state and operations.
The state of $DT^e$ includes both dynamic and communication components:
The dynamic component comprises position $q$, velocity $v$, and acceleration $a_{cc}$, while the communication component includes connection conditions $\eta_{nec}$ and quality-of-service (QoS) requirements $Q_S$. 
Thus, the state of $DT^e$ is represented as the tuple $[q,v,a_{cc},\eta_{nec},\eta_{QoS}]$. 
Moreover, the operation of $DT^e$ involves adjusting the eVTOL’s dynamic status through potential field hyperparameters: $\{k_{tar}, k_{sep}, k_{com}\}$ where $k_{tar}$, $k_{sep}$, and $k_{com}$ are the target field hyperparameter, separation field hyperparameter, and communication field hyperparameter, respectively.
The lifecycle of $DT^e$ simulates eVTOL flight through a high-fidelity virtual environment, enabling accurate aviation modeling within prescribed air corridors.

Similarly, the SIM Digital Twin ($DT^S$) mirrors the behavior of the physical SIM antenna by replicating its phase shift matrix $\bm{\Psi}$ and transmission power allocation $\bm{P}$. The primary function of $DT^S$ is to optimize transmission power and phase shifts, enhancing the air-ground communication capacity between eVTOLs and the ATCo station.

\begin{figure}[h]
\centering
     \includegraphics[width=.49\textwidth]{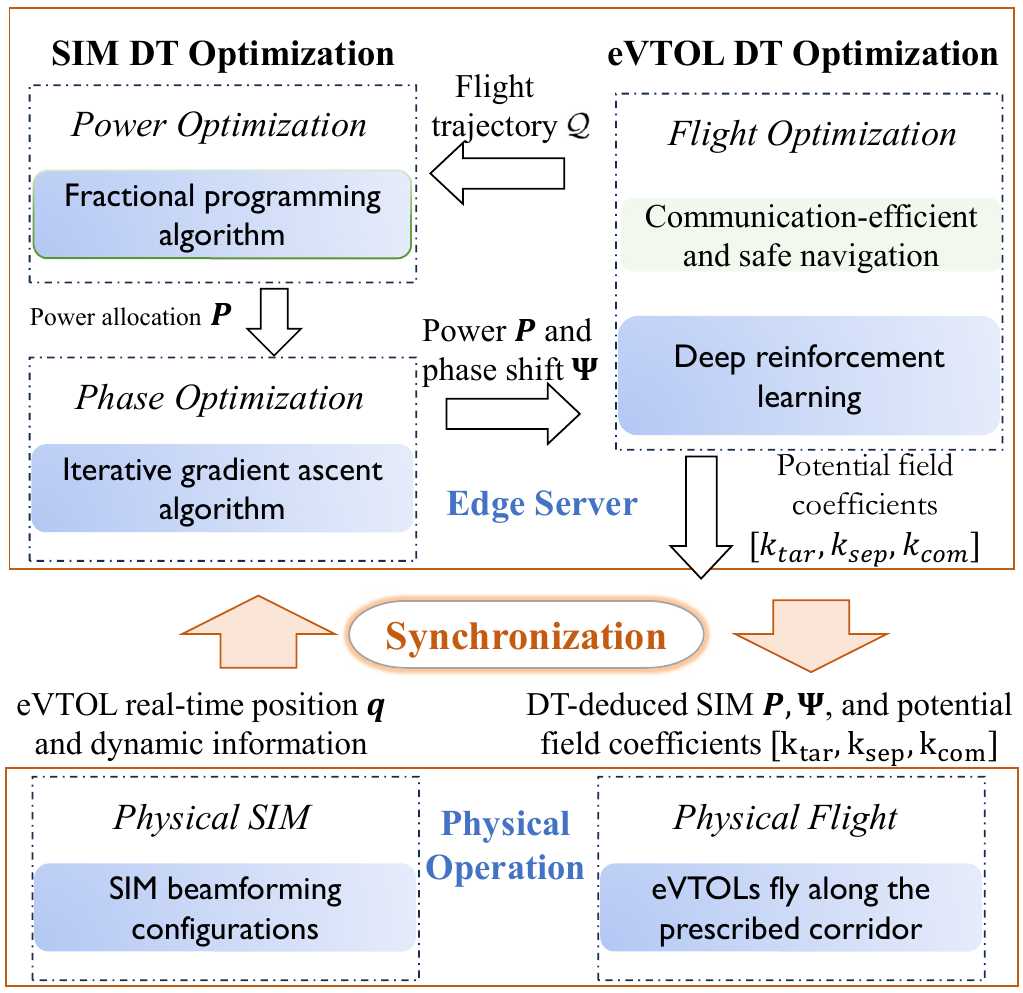} 
     \caption{Flowchart for joint optimization of communication and flight in DT synchronization.} 
\label{sys_flowchart}
\end{figure}

As illustrated in Fig.~\ref{sys_flowchart}, the joint optimization process alternates between $DT^S$ and $DT^e$, both operating on the edge server at the ATCo station.

\textbf{Optimization by $DT^S$}: The alternative DT optimization begins with $DT^S$, which performs a two-step optimization:
The transmission power $P[n]$ is updated using a fractional programming algorithm.
The phase shift matrix $\Psi[n]$ is optimized through an iterative gradient ascent algorithm.
These optimized $DT^S$ parameters are then sent to $DT^e$ for flight control adjustments.

\textbf{Optimization by $DT^e$}: Based on the SIM configurations deduced by $DT^S$, $DT^e$ computes the optimal potential field hyperparameters $\{k_{tar}, k_{sep}, k_{com}\}$. Using these hyperparameters, $DT^e$ yields the eVTOL flight trajectory $\mathcal{Q}[n]$.
 
\textbf{Feedback Loop}:
The output trajectory $\mathcal{Q}[n]$ from $DT^e$ is used as input for the next iteration of SIM optimization in $DT^S$. Additionally, the optimal potential field hyperparameters from $DT^e$ are transmitted to the physical eVTOL for real-time flight control updates.
Further, the physical eVTOL will regularly provide feedback on its current dynamic information to the ATCo station to correct DT trajectory $\mathcal{Q}[n]$ bias. This revised trajectory will in turn alter the SIM optimization of $DT^S$, and so on.

After several iterations between $DT^S$ and $DT^e$, the ATCo station finalizes the optimized parameters, including:
transmission power $P[n]$,
Phase shift matrix $\Psi[n]$,
Potential field hyperparameters $\{k_{tar}, k_{sep}, k_{com}\}$.
These parameters are then synchronized to the physical SIM antenna and eVTOLs. On the physical side, the SIM parameters refine the beamforming of the physical SIM, while the field hyperparameters determine the physical eVTOL flight along the prescribed air corridor.

To maintain synchronization between the physical systems and their DT counterparts, eVTOLs regularly report their current kinetic and communication statuses to the ATCo station. This continuous data exchange ensures that the DT models remain accurate and capable of correcting any DT deduction biases.
The accuracy of DT deductions depends heavily on the frequency of synchronization between the edge server at the ATCo station and the physical eVTOLs. Higher synchronization frequencies improve alignment between the DT and real-world operations, ensuring safe and efficient aviation.

\subsection{Communication and Flight Joint Optimization}
This section formulates the joint optimization model for SIM communication and flight control. The position of the ATCo station is represented as $B=\left[x^{ATC}, y^{ATC}, 0\right]^T$, where $x^{ATC}$ and $y^{ATC}$ denote the station’s x- and y-coordinates, respectively. The z-coordinate is set to $0$, indicating that the ATCo station is located on the ground.

\textbf{eVTOL Flight Model}:
eVTOLs fly within a prescribed air corridor over a total of $N$ time slots, each lasting $\delta$. The position of the $m$-th eVTOL at time slot $n$ is denoted as $q_{m}[n]=\left[x_{m}^{eVTOL}[n], y_{m}^{eVTOL}[n], z_{m}^{eVTOL}[n]\right]^T$, where $m$ is the eVTOL index. The air corridor region is defined as $\mathcal{R}_{cor}$. The eVTOL position $q_{m}[n]$ should satisfy the following constraints:

\begin{equation}
\begin{gathered}
\|q_{m}[n]-q_{m}[n-1]\| \leq V_{\max } \delta, \forall n \in N,\forall m \in M, \\
\end{gathered}
\end{equation}
\begin{equation}
\begin{gathered}
q_{m}[n] \in \mathcal{R}_{cor}, \forall n \in N,\forall m \in M,
\end{gathered}
\end{equation} 

\noindent where $V_{\max }$ is the maximum flight velocity, and the eVTOL must remain within the air corridor $q_{m}[n] \in \mathcal{R}_{cor}$ as required by air traffic regulations. The initial and destination positions of the $m$-th eVTOL are set as:
\begin{equation}
\begin{gathered}
\ q_{m}[0]=f_{m}[0],q_{m}[N]=f_{m}[N],\forall m \in M,
\end{gathered}
\end{equation}

\noindent where $f_{m}[0]$ and $f_{m}[N]$ represent the entry and exit positions of the air corridor.

\textbf{SIM Communication Model}: The transmission power of the SIM antenna is subject to the following constraint:
\begin{equation}
\begin{aligned}
\sum_{m-1}^{M}p_{m}\bigl[n\bigr]\leq P_{ATC},n\in N,
\end{aligned}
\end{equation}

\noindent where $p_{m}[n]$ is the transmission power allocated to the $m$-th eVTOL at time slot $n$. $P_{ATC}$ is the total available transmission power at the ATCo station. Each antenna serves one eVTOL for data transmission, implying that the number of antennas equals $M$.

Next, we specify the physical structure of the SIM. The SIM comprises $L$ metasurface layers, each containing $K$ meta-atoms. 
The phase shift of the $k$-th meta-atom on the $l$-th metasurface layer at time slot $n$ is expressed as $e^{j\theta_{k}^{\l}[n]}$. Consequently, the phase shift matrix for layer $l$ is represented as $\Psi^{l}\Big[n\Big]=diag\Big(e^{j\theta_{1}^{\l}[n]},e^{j\theta_{2}^{\l}[n]},\cdots,e^{j\theta_{K}^{\l}[n]}\Big)\in C^{K\times K},l\in L,n\in N$. 
The transmission matrix $W^{l}\in C^{K\times K},\forall l\neq1,l\in L$ models signal transmission between metasurface layers. The transmission vector from the ATCo station’s antenna to the first metasurface layer is written as $w_{m}^{1}\in C^{K\times 1}$.

According to the Rayleigh-Sommerfeld diffraction theory\cite{{Jiancheng10158690},{SFJGE3F874BA4D351D440F74F99AF523D488}}, the $(k,k^{\prime})$-th entry $w_{k,k^{\prime}}^{l}$ of $W^{l}$ is given by,
\begin{equation}
w_{k,k^{\prime}}^{l}=\frac{d_{x}d_{y}\cos\chi_{k,k^{\prime}}^{l}}{d_{k,k^{\prime}}^{l}}\Bigg(\frac{1}{2\pi d_{k,k^{\prime}}^{l}}-j \frac{1}{\lambda}\Bigg)e^{j2\pi d_{k,k^{\prime}}^{\prime}/\lambda},
\label{uifhsgaf}
\end{equation} 

\noindent where $\lambda$ is the wavelength.
$d_{k,k^{\prime}}^{l}$ is the distance between the $k$-th meta-atom of layer $(l-1)$ and the $k^{\prime}$-th meta-atom of layer $l$.
$\chi_{k,k^{\prime}}^{l}$ is the angle between the propagation direction and the normal to layer $(l-1)$.
$d_{x}\times d_{y}$ is the size of each meta-atom.

Similarly, the $k$-th entry $w_{k,m}^{l}$ of $w_{m}^{1}$ is also derived from Eq.~(\ref{uifhsgaf}). The SIM beamforming matrix $G[n]$ is then given by,
\begin{equation}
G[n]=\Psi^{L}[n]W^{L}\Psi^{L-1}[n]\cdots\Psi^{2}[n]W^{2}\Psi^{1}[n]\in C^{K\times K}.
\label{akugyfiuebvausd}
\end{equation}

\noindent It is worth noting that hardware imperfections may cause deviations from the theoretical model, though these can typically be calibrated before deployment \cite{2022A}.

The baseband equivalent channel from the last metasurface layer of the SIM to the $m$-th eVTOL at time slot $n$, denoted as $h_{m}^{H}[n]\in C^{1\times K}$, is modeled as a Rician channel. The $k$-th entry of $h_{m,k}[n]$ is:
\begin{equation}
h_{m,k}[n]=\sqrt{\frac{\rho_{0}}{(d_{m}[n])^{\alpha^{h}}}}\sqrt{\frac{\kappa^{h}}{\kappa^{h}+1}}\bar{h}_{m}[n],
\label{asdfgeruikyfgdvbxckajhgveariakudsygf}
\end{equation}

\noindent where $\rho_{0}$ is the reference path loss at a distance of $1$ meter, $\kappa^{h}$ is the Rician factor, and $\alpha^{h}$ is the path loss exponent for air-ground communication between the ATCo station and eVTOLs. 
$\bar{h}_{m}[n]=1$ is assumed without loss of generality. The distance $d_{m}[n]$ from the ATCo station to the $m$-th eVTOL at time slot $n$ is:
The distance $d_{m}[n]$ between the ATCo station and the $m$-th eVTOL at time slot $n$ is given by:
\begin{equation}
\begin{aligned}
&d_{m}\left[n\right] =\left\|q_{m}\left[n\right]-B\right\|\\
&=\sqrt{\left(x_{m}^{eVTOL}\left[n\right]-x^{ATC}\right)^{2}+\dots+\left(z_{m}^{eVTOL}\left[n\right]-0\right)^{2}},
\end{aligned} 
\label{rewavcrtfdcred}
\end{equation}

The transmission data for the $m$-th eVTOL in time slot $n$, denoted by $s_{m}[n]$, is assumed to be independent and identically distributed (i.i.d.) with zero mean and unit variance.
The composite signal $y_{m}[n]$ received by the $m$-th eVTOL at time slot $n$, after propagating through the SIM beamforming, is:
\begin{equation}
y_{m}[n]=h_{m}^{H}[n]G[n]\sum_{m=1}^{M}w_{m}^{1}\sqrt{p_{m}[n]}s_{m}[n]+\tau,
\label{asdhgfhaeruikydvfbxckljzuhfiau}
\end{equation}

\noindent where $\tau\sim CN(0, \sigma_{m}^{2})$ represents the i.i.d. additive white Gaussian noise (AWGN) with receiver noise power $\sigma_m^2$ at the eVTOL. Thus, the signal-to-interference-plus-noise ratio (SINR) for the $m$-th eVTOL at time slot $n$ is,
\begin{equation}
\textit{SINR}_{m}[n]=\frac{\left|h_m^H[n] G[n] w_m^1\right|^2 p_m[n]}{\sum_{m^{\prime}=1, m^{\prime} \neq m}^M\left|h_{m^{\prime}}^H[n] G[n] w_{m^{\prime}}^1\right|^2 p_{m^{\prime}}[n]+\sigma_m^2}
\end{equation}

\noindent Accordingly, the data rate of the $m$-th eVTOL at time slot $n$ is,
\begin{equation}
R_{m}[n] = \log\left(1+\textit{SINR}_{m}[n]\right).
\label{qergyuqeruaivgugfkjqwehehFSADCzk}
\end{equation}


\subsection{Problem Formulation}
This section outlines the optimization model to maximize the sum rate of all eVTOLs by jointly optimizing SIM transmission power allocation, phase shift control, and eVTOL flight control. 
The optimization problem is formulated as,
\begin{equation}
\begin{aligned}
P1: &\max_{\bm{P}, \bm{\Psi},\mathcal{Q}} \ \ g(\bm{P}, \bm{\Psi},\mathcal{Q}) = \sum_{n=1}^{N}\sum_{m=1}^{M}R_{m}[n] \\
\text{s.t.} \ \ 
&C1:\sum_{m=1}^{M}p_{m}[n]\leq P_{ATC},  \forall n \in N \\
&C2:p_{m}[n]\geq0, \forall n \in N, \forall m \in M \\
&C3:\theta_{k}^{l}[n]\in[0,2\pi), \forall n\in N,\forall k\in K,\forall l\in L \\
&C4:\|q_{m}[n]-q_{m}[n-1]\|\leq V_{\mathrm{max}} \delta, \forall n \in N, \forall m \in M \\
&C5:q_{m}[n] \in \mathcal{R}_{cor}, \forall n \in N, \forall m \in M \\
&C6:q_{m}[0]=f_{m}[0],q_{m}[N]=f_{m}[N], \forall m \in M ,
\end{aligned}
\label{cargjhvfjdscadsckgvsdajhg}
\end{equation}

\noindent Here, $p_{m}[n]$ denotes the transmission power allocated to the $m$-th eVTOL at time slot $n$. The vector $p_{m}\underline{\underline\Delta}[p_{m}[1],p_{m}[2],\cdots, p_{m}[N]]^{T}$ represents the power allocation for the $m$-th eVTOL across all time slots. The set of all power allocations for the eVTOLs is $\bm{P}\underline{\underline\Delta}\{p_{1},p_{2},\cdots,p_{M}\}$. 
$\psi_{k}^{l}\underline{\underline\Delta}[\theta_{k}^{l}[1],\theta_{k}^{l}[2],\cdots,\theta_{k}^{l}[N]]^{\mathrm{T}}$ represents the phase shift of the $k$-th meta-atom on the $l$-th metasurface layer across all time slots. The phase shift for the $l$-th metasurface layer is $\Psi^{l}\underline{\underline\Delta}\{\psi_{1}^{l},\psi_{2}^{l},\cdots,\psi_{K}^{l}\}$. The combined phase shift matrices for all metasurface layers are denoted as $\bm{\Psi}{\underline{\underline\Delta}}\{{\Psi^{1}},{\Psi^{2}},\cdots,{\Psi^{L}}\}$.
In addition, $q_{m}\underline{\underline\Delta}[q_{m}[0],q_{m}[2],\cdots, q_{m}[N]]^{T}$ represents the trajectory of the $m$-th eVTOL over the time slots. The set of all eVTOL trajectories is $\mathcal{Q}\underline{\underline\Delta}[q_{0},q_{1},\cdots, q_{M}]^{T}$.

The constraints are defined as follows:
C1: The total transmission power must not exceed $P_{ATC}$ in any time slot.
C2: transmission power must remain non-negative.
C3: Phase shifts must lie within the range $[0,2\pi)$.
C4: The eVTOL velocity must not exceed $V_{max}$, given the time slot duration $\delta$.
C5: eVTOLs must stay within the prescribed air corridor $\mathcal{R}_{cor}$, determined by the corridor's shape.
C6: Initial and final positions of each eVTOL are fixed at $f_m[0]$ (entry) and $f_m[N]$ (exit).

The problem $P1$ involves three interdependent optimization variables:
transmission power allocation ($P$), SIM phase shift matrix ($\bm{\Psi}$), and eVTOL flight trajectory ($\mathcal{Q}$). Due to the interdependence of these variables, $P1$ is a non-convex problem that requires iterative methods to achieve a solution. The next section introduces an iterative algorithm to obtain sub-optimal results for $P1$.

\section{Model Solution}
This section presents a solution strategy using the Block Coordinate Descent (BCD) method \cite{2021Block} to alternately optimize the variables in $P1$. The problem is divided into three blocks, corresponding to $\bm{P}$, $\bm{\Psi}$, and $\mathcal{Q}$
Each block is optimized sequentially, keeping the other two blocks fixed during each step.

The proposed BCD algorithm integrates the dual DT design ($DT^S$ and $DT^e$). The workflow is as follows:
Optimize transmission power ($\bm{P}$): Within $DT^S$, given fixed  $\bm{\Psi}$ and $\mathcal{Q}$, the transmission power allocation $\bm{P}$ is updated using an approximate linear method, as described in \cite{2018Fractional}. Update Phase Shift Matrix ($\bm{\Psi}$): Using the updated $\bm{P}$ and fixed $\mathcal{Q}$, $DT^S$ applies an iterative gradient ascent algorithm to optimize the phase shift matrix $\bm{\Psi}$. Refine Flight Trajectory ($\mathcal{Q}$): With updated $\bm{P}$ and $\bm{\Psi}$, $DT^e$ optimizes the eVTOL trajectories $\mathcal{Q}$ using a composite potential field method.
After several BCD iterations, the optimized parameters ($\bm{P}$, $\bm{\Psi}$, $\mathcal{Q}$) guide both SIM beamforming and eVTOL navigation. This iterative approach improves the efficiency of communication and flight within prescribed air corridors. In the following subsections, we detail the optimization process for each block.


    
    



        

\subsection{SIM Communication Optimization}
The optimization for $DT^S$ involves two components: the transmission power $\bm{P}$-related block and the phase shift matrix $\bm{\Psi}$-related block.

\subsubsection{transmission power $\bm{P}$-related block optimization}
For the transmission power block, with the SIM phase shift matrix $\bm{\Psi}$ and the eVTOL trajectory $\mathcal{Q}$ fixed, the optimization problem reduces to $P(1a)$,
\begin{equation}
\begin{aligned}
P(1a): &\max g_{a}(\bm{P})=\sum_{n=1}^{N}\sum_{m=1}^{M}R_{m}[n] \\
\text{s.t.}  \ \ &C1:\sum_{m-1}^{M}p_{m}[n]\leq P_{ATC},\forall n\in N \\
&C2: p_{m}[n]\geq0,\forall n\in N,\forall m\in M.
\end{aligned}
\label{asdjfhvkuyfvgfbdkjcvbuas}
\end{equation}

\noindent This is a typical fractional programming (FP) problem. It can be addressed using the approximate linear method. By employing a Lagrangian dual transformation \cite{2018Fractional}, $g_{a}(\bm{P})$ is reformulated as $g_{a 1}(P, \mu)$,
\begin{equation}
\begin{aligned}
& g_{a 1}(\bm{P}, \mu)=\sum_{n=1}^N \sum_{m=1}^M \log \left(1+\mu_m[n]\right)-\sum_{n=1}^N \sum_{m=1}^M \mu_m[n]+ \\
& \sum_{n=1}^N \sum_{m=1}^M \frac{(1+\mu_m[n])|h_m^H[n] G[n] w_m^1|^2 p_m[n]}{\sum_{m^{\prime}=1}^M\left|h_{m^{\prime}}^H[n] G[n] w_{m^{\prime}}^1\right|^2 p_{m^{\prime}}[n]+\sigma_m^2}.
\label{asdfhjaewgrfkg}
\end{aligned}
\end{equation}

\noindent Here, $\mu = \left[\mu_1, \ldots, \mu_m, \ldots, \mu_M\right]^T$ is an auxiliary variable where $\mu_m[n] \geq 0$ for all $m$ and $n$.
To solve the multi-ratio FP problem in the above equation, the quadratic transformation \cite{2018Fractional} is applied. This converts the problem into a biconvex optimization problem:
\begin{equation}
\begin{aligned}
& g_{a 2}(P, \mu, y)=\sum_{n=1}^N \sum_{m=1}^M\left(\log \left(1+\mu_m[n]\right)-\mu_m[n]\right)+ \\
& \sum_{n=1}^N \sum_{m=1}^M 2 y_m[n] \sqrt{\left(1+\mu_m[n]\right)\left|h_m^H[n] G[n] w_m^1\right|^2 p_m[n]}- \\
& \sum_{n=1}^N \sum_{m=1}^M y_m^2[n]\left(\sum_{m^{\prime}=1}^M\left|h_{m^{\prime}}^H[n] G[n] w_{m^{\prime}}^1\right|^2 p_{m^{\prime}}[n]+\sigma_m^2\right),
\label{etguyhertughyvd}
\end{aligned}
\end{equation}

\noindent where $y=\left[y_{1}, \ldots, y_{m} \ldots, y_{M}\right]^T$ are auxiliary variables. The problem $P(1a)$ can now be rewritten as:
\begin{equation}
\begin{aligned}
 \overline{P(1a)}: &\max g_{a 2}(P, \mu, y) \\
 \text { s.t. } \ \
&C1: \sum_{m=1}^M p_m[n] \leq P_{ATC}, \forall n \in N \\
&C2: p_m[n] \geq 0, \forall n \in N, \forall m \in M \\
&C3: \mu_m[n] \geq 0, \forall n \in N, \forall m \in M
\end{aligned}
\label{yuirewyfgyrfgsdhjg}
\end{equation}

In this reformulated problem $\overline{P(1a)}$, $\bm{P}$, $\mu$, and $y$ are updated iteratively. By fixing $\bm{P}$ and $y$ as constants, the partial derivative of $g_{a2}$ with respect to $\mu_m[n]$ is computed as:

\begin{equation}
\frac{\partial g_{a 2}}{\partial \mu_m[n]}=\frac{1}{1+\mu_m[n]}-1+\frac{y_m[n] \sqrt{\left|\mathcal{H}^\mathcal{G}_m\right|^2 p_m[n]}}{\sqrt{1+\mu_m[n]}},
\label{jdfsgkseuryfgkaugef}
\end{equation}

\noindent where $\mathcal{H}^\mathcal{G}_m = h_m^H[n] G[n] w_m^1$. Setting ${\partial g_{A2}}/{\partial \mu_m[n]} = 0$ gives the optimal value:
\begin{equation}
\begin{aligned}
&\mu^*_m[n]=\frac{y_m^2[n]\left|\mathcal{H}^\mathcal{G}_m\right|^2 p_m[n]}{2}+ \\
&\frac{y_m[n] \sqrt{\left|\mathcal{H}^\mathcal{G}_m\right|^2 p_m[n] (y_m^2[n]\left|\mathcal{H}^\mathcal{G}_m\right|^2 p_m[n]+4)}}{2}. \\
\end{aligned}
\label{dsahfjgaerdfvytguasiuyweio}
\end{equation}

\noindent Similarly, setting ${\partial g_{a2}}/{\partial y_m[n]} = 0$ gives the optimal $y^*_m[n]$:
\begin{equation}
y^*_m[n]=\frac{\sqrt{\left(1+\mu_m[n]\right)\left|h_m^H[n] G[n] w_m^1\right|^2 p_m[n]}}{\sum_{m^{\prime}=1}^M\left|h_{m^{\prime}}^H[n] G[n] w_{m^{\prime}}^1\right|^2 p_{m^{\prime}}[n]+\sigma_m^2}.
\label{sdaiuotrifaesdjkvcb}
\end{equation}

Next, the dual function of $g_{a2}$ is introduced by incorporating dual variables $\beta[n]$ and $a_m[n]$ to account for SIM transmission power constraints:
\begin{equation}
\begin{aligned}
& g_{a 2}^{Dual}(P, \beta, a)=\sum_{n=1}^N \sum_{m=1}^M\left(\log \left(1+\mu_m[n]\right)-\mu_m[n]\right)+ \\
& \sum_{n=1}^N \sum_{m=1}^M 2 y_m[n] \sqrt{\left(1+\mu_m[n]\right)\left|\mathcal{H}^\mathcal{G}_m\right|^2 p_m[n]}- \\
& \sum_{n=1}^N \sum_{m=1}^M y_m^2[n]\left(\sum_{m^{\prime}=1}^M\left|h_{m^{\prime}}^H[n] G[n] w_{m^{\prime}}^1\right|^2 p_{m^{\prime}}[n]+\sigma_m^2\right)+ \\
& \sum_{n=1}^N \beta[n]\left(\sum_{m=1}^M p_m[n]-P_{ATC}\right)+a_m[n]\left(-p_m[n]\right).
\end{aligned}
\end{equation}

\noindent Using the Karush-Kuhn-Tucker (KKT) conditions \cite{ghojogh2021}, the optimal transmission power $\bm{P}^*$ can be derived by solving the system of equations:
\begin{equation}
\left\{\begin{array}{l}
\frac{\partial g_{a 2}^{Dual}}{\partial p_m[n]}=\frac{y_m[n] \sqrt{\left(1+\mu_m[n]\right)\left|h_m^H[n] G[n] w_m^1\right|^2}}{\sqrt{p_m[n]}}- \\
\left|h_m^H[n] G[n] w_m^1\right|^2 \sum_{m=1}^M y_m^2[n]+\beta[n]- \\
a_m[n]=0, \forall n \in N, \forall m \in M \\
\sum_{m=1}^M p_m[n] \leq P_{ATC}, \forall n \in N \\
\beta[n] \geq 0, \forall n \in N \\
\beta[n]\left(\sum_{m=1}^M p_m[n]-P_{ATC}\right)=0, \forall n \in N \\
p_m[n] \geq 0, \forall n \in N, \forall m \in M \\
a_m[n] \geq 0, \forall n \in N, \forall m \in M \\
-a_m[n] p_m[n]=0, \forall n \in N, \forall m \in M
\end{array}\right.
\label{adsfvjhgardsfdfkgfery}
\end{equation}

The solving technique is as follows: setting $a_{m}[n]=0, \forall n \in N, \forall m \in M$ and $\frac{\partial g_{a 2}^{Dual}}{\partial p_m[n]}=0, \forall n \in N, \forall m \in M$, we get,
\begin{equation}
p_m[n]=\left(\frac{y_m[n] \sqrt{\left(1+\mu_m[n]\right)\left|h_m^H[n] G[n] w_m^1\right|^2}}{\left|h_m^H[n] G[n] w_m^1\right|^2 \sum_{m=1}^M y_m^2[n]-\beta[n]}\right)^2
\label{pmn}
\end{equation}

\noindent Here, $\beta[n]$ is gradually increased from 0 until the condition $0.9 P_{ATC} \leq \sum_{m=1}^M p_m[n] \leq P_{ATC}, \forall n \in N$ is satisfied. An exponential increase and linear decrease search algorithm is applied to obtain the suboptimal solution $\bm{P}^*$. The complete iterative process is summarized in Alg.~\ref{Ag2}.

\begin{algorithm} 
    \caption{Prox-Linear Algorithm for $P(1a)$} \label{Ag2}

   Initialize {$\bm{P}^0$, $y^0$} and $t^a=0$.
   
\While{$t^a\neq t_{max}^a$} {

    Update $\mu$ using Eq.~(\ref{dsahfjgaerdfvytguasiuyweio});

    Update $y$ using Eq.~(\ref{sdaiuotrifaesdjkvcb});

    Update $\bm{P}^*$ by solving Eq.~(\ref{adsfvjhgardsfdfkgfery});

    $t^a=t^a+1$.
}
\end{algorithm}

\subsubsection{Phase Shift Matrix $\bm{\Psi}$-related block optimization}

To optimize the SIM phase shift matrix $\bm{\Psi}$, the original optimization problem $P1$ is reduced to $P(1b)$, assuming that the SIM transmission power $\bm{P}$ and eVTOL flight trajectory $\mathcal{Q}$ are fixed:
\begin{equation}
\begin{aligned}
P(1b):&\max g_{b}(\bm{\Psi})=\sum_{n=1}^{N}\sum_{m=1}^{M}R_{m}[n] \\
\text{s.t.} \ \ &\theta_{k}^{l}[n]\in[0,2\pi),\forall n\in N,\forall k\in K,\forall l\in L
\end{aligned}
\label{sdahfgiuubrituoehier}
\end{equation}

To solve $P(1b)$, a gradient ascent algorithm is applied. Initially, the phase shifts $\theta_{k}^{l}[n]$ are randomly initialized using a uniform distribution. The partial derivative of $g_b(\bm{\Psi})$ with respect to $\theta_k^l[n]$ is calculated as:
\begin{equation}
\frac{\partial g_b(\bm{\Psi})}{\partial \theta_k^l[n]}=\sum_{m=1}^M \gamma_m p_m[n] \eta_m[n]
\end{equation}
where $\gamma_m$ and $\eta_m[n]$ are given by
\begin{equation}
\gamma_m=\frac{1}{\sum_{m^{\prime}=1}^M\left|h_{m^{\prime}}^H[n] G[n] w_{m^{\prime}}^1\right|^2 p_{m^{\prime}}[n]+\sigma_m^2},
\end{equation}

\noindent and
\begin{equation}
\begin{aligned}
& \eta_m[n]=2 \times \operatorname{Im}\left\{h_m^H[n] G[n] w_m^1\right\} \times \operatorname{Im}\left(\varsigma_{k, l, n, m}\right)+ \\
& 2 \times \operatorname{Re}\left\{h_m^H[n] G[n] w_m^1\right\} \times \operatorname{Re}\left(\varsigma_{k, l, n, m}\right), \\
\end{aligned}
\label{hjgkjskrgudfhbu}
\end{equation}

\noindent respectively, in which $\varsigma_{k, l, n, m}$ is,
\begin{equation}
\begin{aligned}
\varsigma_{k, l, n, m}=h_m^H[n] v_k^l[n] e^{j \theta_k^l[n]} u_k^l[n] w_m^1 j. 
\end{aligned}
\label{fbsghiujhsgiusrtgiu}
\end{equation}

\noindent In this equation, $u_k^l[n]$ and $v_k^l[n]$ represent the $k$-th row and $k$-th column of the matrices $U_l[n] \in C^{K \times K}$ and $V_l[n] \in C^{K \times K}$, respectively. These matrices are defined as:
\begin{equation}
\begin{aligned}
& U^l[n] \stackrel{\Delta}{=}\left\{\begin{array}{l}
W^l \Psi^{l-1}[n] \cdots \Psi^2[n] W^2 \Psi^1[n], \text if\ l \neq 1, \\
I_K, if\ l=1,
\end{array}\right. \\
\end{aligned}
\end{equation}
\begin{equation}
\begin{aligned}
& V^l[n] \stackrel{\Delta}{=}\left\{\begin{array}{l}
\Psi^L[n] W^L \Psi^{L-1}[n] \cdots \Psi^{l+1}[n] W^{l+1}, if\ l \neq L, \\
I_K, if\ l=L.
\end{array}\right.
\label{yhgbuyrtugiydsoiaufyu}
\end{aligned}
\end{equation}

\begin{algorithm} 
    \caption{Iterative Gradient Ascent for $P(1b)$} \label{Ag3}
   Initialize {$\bm{\Psi}[0]$}, compute $\nabla g_b\left(\bm{\Psi}[0]\right)$ and set $t^b=1$, $f=0.001$, $step=0.5$, and $\epsilon_{thr}^{b}$.

\While {$g_b(\bm{\Psi}[t^{b}])-g_b(\bm{\Psi}[t^{b}-1]) > \epsilon_{thr}^{b}$} {
    $count=1$\\
    \While{$g_b\left(\bm{\Psi}[t^b]+count \times \nabla g_b\left(\bm{\Psi}[t^b]\right)\right)<g_b\left(\bm{\Psi}[t^b]\right)+f \times count \times \nabla g_b\left(\bm{\Psi}[t^b]\right)^T \nabla g_b\left(\bm{\Psi}[t^b]\right)$:}{
    $count = count \times step$;}
    According to Eq.~(\ref{uioyuoierwugdfg}), \\$\bm{\Psi}[t^b+1]=\bm{\Psi}[t^b]+count \times \nabla g_b\left(\bm{\Psi}[t^b]\right)$ \\
    $t^b = t^b + 1$ \\
}
\end{algorithm}

Once obtain all the partial derivatives $\frac{\partial g_b(\bm{\Psi})}{\partial \theta_k^{l}[n]}$, the phase shifts $\{\theta_k^{l}[n]\}$ are updated iteratively:
\begin{equation}
\theta_{k}^{l}[n+1]\leftarrow\theta_{k}^{l}[n]+\xi\frac{\partial g_{b}(\bm{\Psi})}{\partial\theta_{k}^{l}[n]},\forall n\in N,\forall k\in K,\forall l\in L
\label{uioyuoierwugdfg}
\end{equation}
where $\xi > 0$ is the step size, determined using the Armijo backtracking line search \cite{2021Intelligent}. 
The process repeats until the transmission rate improvement between consecutive iterations falls below the threshold $\epsilon_{thr}^b$.
Finally, we get the sub-optimal phase shift matrix $\bm{\Psi}$.
The iterative gradient ascent algorithm for optimizing $P(1b)$ is outlined in Alg.~\ref{Ag3}.


\subsection{Flight Control Optimization}
This section addresses the optimization of eVTOL trajectories to maximize communication transmission rates while ensuring safe aviation along a prescribed corridor. The trajectory optimization is achieved using the proposed Composite Potential Field (CPF) algorithm, which incorporates three potential fields: target field $\mathcal{F}^{tar}$, separation field $\mathcal{F}^{sep}$, and communication field $\mathcal{F}^{com}$. Consequently, the trajectory optimization problem for $\mathcal{Q}$ is reformulated into an optimization problem with respect to the potential field hyperparameters $\{k_{tar}, k_{sep}, k_{com}\}$, where $k_{x}$ is the field hyperparameter corresponding to each potential field $\mathcal{F}^{x}$.

Given fixed SIM transmission power allocation $\bm{P}$ and phase shift matrix $\bm{\Psi}$, the original problem $P1$ is reduced to $P(1c)$, as defined below:
\begin{equation}
\begin{aligned}
P(1c)&: \max_{k_{tar}, k_{sep}, k_{com}}  {g}_c(\mathcal{Q})= \sum_{m=1}^{M} \left\{  \mathcal{S}_{m} + \sum_{n=1}^{N} R_{m}[n] \right\}  \\
\text{s.t.}  \ \ &C1:\|q_{m}[n]-q_{m}[n-1]\|\leq V_{\mathrm{max}} \delta, \forall n \in N, \forall m \in M \\
&C2:q_{m}[n] \in \mathcal{R}_{cor}, \forall n \in N, \forall m \in M \\
&C3:q_{m}[0]=f_{m}[0],q_{m}[N]=f_{m}[N], \forall m \in M ,
\end{aligned}
\label{asdjfhvkuyfvgfbdkjcvbuas}
\end{equation}

\noindent Here, $\mathcal{S}_{m} = \sum_{k=1,k\neq m}^{M}\frac{1-e^{c_1+c_2d_{k,m}}}{M}$, where $c_1$ and $c_2$ are weighting constants, and $d_{k,m}$ is the distance between eVTOLs $k$ and $m$. 

The optimization objective of $P(1c)$ consists of two key parts:
(i) \textit{Collision Risk Minimization}:
The term $\mathcal{S}_m$ penalizes potential collisions between eVTOLs. A significant penalty is applied when the distance $d_{k,m}$ between eVTOLs falls too small, determined by constants $c_1$ and $c_2$.
(ii) \textit{Communication Rate Maximization}: 
This term $\sum_{n=1}^{N} R_{m}[n]$ rewards the total SIM communication rate between the ATCo station and the eVTOLs.
To address this optimization problem, the CPF method is employed. This method extracts and adjusts flight control hyperparameters to ensure safe and communication-efficient eVTOL aviation.

\subsubsection{Composite Potential Field}
The CPF approach integrates multiple potential fields to provide flight control solutions that are simple, localized, and capable of distributed calculation \cite{9143127Enming}. The CPF design focuses on two main objectives: safe aviation and communication connectivity. Each potential field is described below:



(i) \textbf{Safe Aviation}: is achieved through the combination of the target field and the separation field:
The target field $\mathcal{F}_{i}^{tar}[n]$ guides eVTOLs toward their intended destinations and is defined as:
\begin{equation}
    \mathcal{F}_{i}^{tar}[n] = \frac{1}{2} k_{tar} \|q_{i}[n] - q_{tar}\|^2,
\label{dsarfguyersgyure}
\end{equation}

\noindent where $k_{tar}$ is the target field hyperparameter, and $\|q_{i}[n] - q_{tar}\|$ represents the distance between eVTOL $i$ and its target position at time slot $n$.

Meanwhile, the separation field $\mathcal{F}_{i}^{sep}[n]$ ensures a safe distance is maintained between eVTOLs, preventing collisions. It is expressed as:
\begin{equation}
 \mathcal{F}_{i}^{sep}[n]= 
 \sum_{j=1,j\neq i,j\in N_i}^{|N_i|}  \mathcal{F}_{i,j}^{sep}[n],
\end{equation}

\noindent where,
\begin{equation}
 \mathcal{F}_{i,j}^{sep}[n]= 
\begin{cases} 
 \frac{1}{2}k_{sep}\left(\frac{d_{sep}}{d_{i,j}[n]}\right)^2,& \text{if } d_{i,j}[n] \leq d_{sep} \\
 0, & \text{if } d_{i,j}[n] > d_{sep}.
 \end{cases} 
\end{equation}

\noindent In this equation, $k_{sep}$ is the separation field hyperparameter, $d_{sep}$ is the minimum safe separation between eVTOLs.
$N_i$ represents the neighboring eVTOLs of $i$.

(ii) \textbf{Communication Connectivity}: The communication field $\mathcal{F}_{i}^{com}[n]$ ensures that eVTOLs remain connected within the platoon. It is defined as:
\begin{equation}
 \mathcal{F}_{i}^{com}[n]= 
 \sum_{j=1,j\neq i,j\in N_{i}^{c}}^{|N_{i}^{c}|}  \mathcal{F}_{i,j}^{com}[n],
 \label{sakjdheygf}
\end{equation}

\noindent where
\begin{equation}
\mathcal{F}_{n,i,j}^{com} = 
\begin{cases} 
    \frac{1}{2}k_{com} d_{i,j}[n]^2 & \text{if} \ \ d^{eVTOL}_{com} \geq d_{i,j}[n] \geq d^{eVTOL}_{max} \\
    0 & \text{otherwise},
\end{cases}
\end{equation}

\noindent Here, $k_{com}$ is the communication field hyperparameter, $d_{i,j}[n]$ represents the distance between eVTOLs $i$ and $j$, and $N_i^c$ is the set of communication-active neighbors. 
The threshold $d^{eVTOL}_{com}$ ensures proper signal reception, while $d^{eVTOL}_{max}$ defines the maximum communication range, preventing eVTOL-to-eVTOL communication disconnection.


Finally, the flight acceleration $a_{i}[n]$ of eVTOL $i$ is derived from the negative gradient of the combined potential fields:
\begin{equation}
a_{i}[n] = -\nabla{(\mathcal{F}_{i}^{com}[n] + \mathcal{F}_{i}^{sep}[n] + \mathcal{F}_{i}^{tar}[n])}.
\label{sdafyterqyfiqehjgkyrafgid}
\end{equation}
This formulation ensures that the eVTOLs adjust their trajectories in real time to balance safety and communication connectivity, achieving optimal eVTOL aviation.

\subsubsection{DQN Framework}
When the field hyperparameters $\{k_{tar}, k_{sep}, k_{com}\}$ remain fixed, the resulting CPF force is constant. This leads to either linear or circular motion in the eVTOL trajectory, which is inadequate for navigating eVTOL through complexly shaped corridors. Additionally, the objective value of $P(1c)$ fluctuates with environmental changes (such as suddenly appearing obstacles: birds, drones), necessitating adaptive field hyperparameters to handle dynamic conditions.

To address this, we propose a Deep Reinforcement Learning (DRL) approach \cite{9410457Chongwen} to dynamically adjust the composite field hyperparameters, ensuring safe and efficient eVTOL flight along prescribed corridors. In the proposed method, the hyperparameters $k_{tar}$, $k_{sep}$, and $k_{com}$ are treated as learnable parameters that are fine-tuned through DRL interactions with the environment.

We employ the Deep Q-Network (DQN) algorithm, a reinforcement learning technique well-suited for high-dimensional state spaces and complex inputs. DQN combines the principles of Q-learning with the representational power of deep neural networks \cite{Tian10195210}.

In general, the DQN framework consists of three fundamental components: State, Action, and Reward.

\textbf{State}:
The state space, denoted as $\mathcal{S}$, encapsulates key environmental information. A state $\mathbf{s} \in \mathcal{S}$ includes the target position $\mathbf{p}_t$ and the relative positions of neighboring eVTOLs. These relative positions are represented as offset values relative to the target position:
\begin{equation}
\mathbf{s} = \{\mathbf{p}_t, (\mathbf{p}_i - \mathbf{p}_t) \mid i \in \mathcal{N}\}
\end{equation}
\noindent where $\mathbf{p}_t$ is the position of the target, $\mathbf{p}_i$ is the position of eVTOL $i$, and $\mathcal{N}$ denotes the set of neighboring eVTOLs.

\textbf{Action}:
The action space, denoted as $\mathcal{A}$, specifies potential adjustments to the field hyperparameters $\{k_{tar}, k_{sep}, k_{com}\}$. For each hyperparameter, changes can be one of three options: \{+0.06, -0.06, or 0 (no change)\}.  Consequently, the action space $\mathcal{A}$ contains $M\times 3\times 3$ possibilities:
\begin{equation}
\begin{split}
\mathcal{A} = &\{\Delta k^i_* \mid \Delta k^i_* \in \{\pm 0.06,  0\}, \, i = 1, \ldots, M, \\ & *\in \{tar,sep,com\}\}
\end{split}
\end{equation}

\textbf{Reward}:
The reward function drives learning by providing feedback for the agent’s actions. For collaborative flight involving a platoon of $M$ eVTOLs, the reward for eVTOL $i$ is defined as:
\begin{equation}
\begin{aligned}
\begin{split}
&r_i = \alpha_1 v_{tar}(1 + \alpha_2 d_{i,tar}) + 
\sum_{ j \neq i, \, d_{i,j} < d_{sep}  }\left(\frac{\beta v_{sep}}{d_{i,j}}\right) \\
 &+ \sum_{m=1}^{M}\log\left(1 + \frac{\left|\mathcal{H}^\mathcal{G}_m \right|^2 p_m[n]}{\sum_{m^\prime=1, m^\prime \neq m}^{K} \left|\mathcal{H}^\mathcal{G}_{m^\prime} \right|^2 p_{m^\prime}[n] + \sigma_m^2} \right)
\end{split}
\end{aligned}
\label{kbuycxvguiyesrjhsdfg}
\end{equation}

\noindent where $v_{tar}$ and $v_{sep}$ are velocity components directing the eVTOL toward the target and away from nearby eVTOLs, respectively. $d_{i,tar}$ denotes the distance between eVTOL $i$ and its destination (target).
$d_{i,j}$ is the distance between eVTOLs $i$ and $j$. 
$\alpha_1$, $\alpha_2$, and $\beta$ are weighting factors that balance the reward components.

\textbf{DQN Network Design}:
A key advantage of DQN is its model-free nature. It learns the optimal action-value function $Q^*(s,a)$ directly from the environment. The Q-function update rule is given by:
\begin{equation}
\begin{aligned}
Q(s[n], a[n]) \leftarrow &(1-\alpha)Q(s[n], a[n]) \\
&+ \alpha \left( r[n+1] + \gamma \max_{a'} Q(s[n+1], a')  \right)
\end{aligned}
\label{kdgjhtriudhidfugvur}
\end{equation}

\noindent where $\alpha$ is the learning rate. 
$r[n+1]$ is the reward at time step $n+1$,
$\gamma$ is the discount factor for future rewards,
$s[n+1]$ represents the next state, and $a'$ is the action that maximizes the Q-value at $s[n+1]$.
To balance exploration and exploitation, an $\epsilon$-greedy policy is used. Here, $\epsilon$ defines the probability of exploring random actions versus exploiting the learned policy.

\begin{figure} [h]
\centering
     \includegraphics[width=0.48\textwidth]{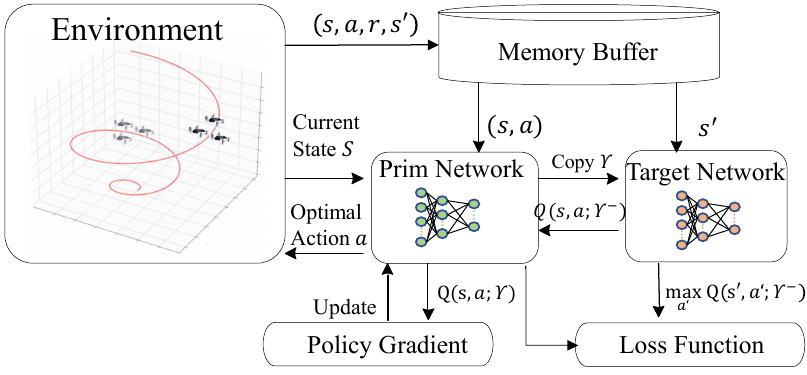} 
     \caption{The illustration of the proposed DQN structure} 
\label{dqn_structure}
\end{figure}
The framework of our DQN network is illustrated in Fig.~\ref{dqn_structure}.
In DQN, the Q-function is approximated using a deep neural network with parameters $\Upsilon$. To stabilize training, a target network $Q_{target}$ with parameters $\Upsilon^-$ is maintained and updated less frequently. The update rule for the target Q-network is:
\begin{equation}
y[n] = r[n] + \gamma \max_{a'} Q_{\text{target}}(s[n+1], a'; \Upsilon^-),
\label{itruierlghrlistugh}
\end{equation}

\noindent where $y[n]$ is the target Q-value. $a'$ denotes the action that maximizes the Q-value at the next state $s[n+1]$. The primary Q-network is trained using the following loss function, i.e., the mean squared error between the predicted and target Q-values:
\begin{equation}
\begin{aligned}
&L(\theta) =\\
&\mathbb{E}_{(s[n], a[n], r[n], s[n+1]) \sim \mathcal{D}} \left[ \left( y[n] - Q_{\text{prim}}(s[n], a[n]; \Upsilon) \right)^2 \right],
\end{aligned}
\end{equation}
\noindent where $\mathcal{D}$ is the experience replay buffer. 
which stores transitions \((s[j], a[j], r[j], s[j+1])\) to reduce the correlation between consecutive experiences and improve training stability.
The network parameters $\Upsilon$ are updated using Stochastic Gradient Descent (SGD):
\begin{equation}
\Upsilon \leftarrow \Upsilon - \eta \nabla_{\Upsilon} L(\Upsilon)
\label{iteiudfjhgk}
\end{equation}

\noindent where $\eta$ is the learning rate.
The loss function for the mini-batch sampled from $\mathcal{D}$ is given by:
\begin{equation}
\begin{aligned}
&L(\Upsilon) = \\
&\mathbb{E}_{(s[j], a[j], r[j], s[j+1]) \sim \mathcal{D}} \left[ \left( y[j] - Q_{\text{prim}}(s[j], a[j]; \Upsilon) \right)^2 \right]
\label{iuhrthvukdsafyvkae}
\end{aligned}
\end{equation}

\noindent where,
\begin{equation}
\begin{aligned}
y[j] = 
\begin{cases} 
r[j]  &\text{Terminal} \\
r[j] + \gamma \max_{a'} Q_{\text{target}}(s[j+1], a'; \Upsilon^-)  &\text{Otherwise}.
\end{cases}
\label{fiutgruthge}
\end{aligned}
\end{equation}

\noindent The proposed DQN framework is illustrated in Fig.~\ref{dqn_structure}.

\begin{algorithm}
\caption{Deep Q-Network (DQN)}
\label{Ag4}
    Initialize Primary network, Target network, and replay memory;
\ForEach{episode = 1 to $M$}{
    \ForEach{t = 1 to $T$}{
        With probability $1-\epsilon$ select $a[n] = \argmax_a Q(s[n], a; \Upsilon)$\;
        Obtain reward $r[n]$ and next state $s[n+1]$\;
        \ForEach{transition $(s[j], a[j], r[j], s[j+1])$ in minibatch}{
            \eIf{$s[j+1]$ is terminal state}{
                $y[j] = r[j]$\;
            }{
                $y[j] = r[j] + \gamma \max_{a'} Q(s[j+1], a'; \Upsilon^-)$\;
            }
        }
        Perform a gradient descent step on Eq.~(\ref{iuhrthvukdsafyvkae});\\
        $s[n] = s[n+1]$\;
        \If{$s[n]$ is terminal}{
            Output the optimal $\{k_{tar}, k_{sep}, k_{com}\}$;\\
            break;
        }
    }
}
\end{algorithm}

As seen in Eq.~(\ref{kbuycxvguiyesrjhsdfg}), the reward function consists of three parts: target proximity reward, obstacle avoidance reward, and SIM transmission rate reward. 
The target proximity reward encourages eVTOLs to reduce the distance to their target destination.
The obstacle avoidance reward penalizes eVTOLs for getting too close to each other, ensuring they maintain a safe separation. Finally, the SIM transmission rate reward incentivizes flight paths that maximize the communication rate between the ATCo station and the eVTOLs. In summary, a higher total reward is achieved when eVTOLs approach their targets, avoid collisions, and improve SIM communication rates.


The typical procedure for DQN is outlined in Alg.~\ref{Ag4}. Each training episode begins by initializing the state $s[n]$ from the environment and iterating for a maximum of $T$ time steps. At each time step, the DQN algorithm selects an action with a probability of $1 - \epsilon$. If the action is not randomly selected, it is determined as $a[n] = \arg\max_a Q(s[n], a; \Upsilon)$, where $\Upsilon$ represents the learnable parameters of the primary network. 
The selected action $a[n]$ is executed, and the environment returns a reward $r[n]$, updating the state to $s[n+1]$. 
The algorithm then minimizes the loss function through a gradient descent step and updates $s[n]$ to $s[n+1]$. 
To ensure training stability, the target network parameters $\Upsilon^-$ are periodically synchronized with the primary network parameters $\Upsilon$. 
The training episode ends when $s[n]$ reaches a terminal state, outputting the optimal field hyperparameters $\{k_{tar}, k_{sep}, k_{com}\}$ for safe and efficient eVTOL flight.




\begin{algorithm} 
    \caption{DT-based Joint Optimization} \label{Ag5}
   Initialize {$\bm{\Psi}, \mathcal{Q}, \bm{P}, \{k_{tar}, k_{sep}, k_{com}\}, \epsilon_{thr}, f_{re}$}\\
   
    \While {$t\le t_{max}$} {
            \While{$|g(\bm{P}[n], \bm{\Psi}[n])-g(\bm{P}[n-1], \bm{\Psi}[n-1])| > \epsilon_{thr}$} {
            
                Fix {$\bm{\Psi}[n]$, $\mathcal{Q}[n]$} and update $\bm{P}[n+1]$ by Alg.~\ref{Ag2}
        
                Fix {$\bm{P}[n+1]$, $\mathcal{Q}[n]$} and update $\bm{\Psi}[n+1]$ by Alg.~\ref{Ag3}
        
                $n = n + 1$
        
            }
            \While{$i \le i_{max}$} {
                Fix {$\bm{P}[i]$, $\bm{\Psi}[i]$} and update $\{k_{tar}[i], k_{sep}[i], k_{com}[i]\}$ by Alg.~\ref{Ag4}\\
                $i = i + 1$
            }
            Update $\mathcal{Q}$ according to $\{k_{tar}, k_{sep}, k_{com}\}$

            Deliver the deduction $\{\bm{P}, \bm{\Psi}, [k_{tar}, k_{sep}, k_{com}]\}$ to the physical SIM antenna and eVTOLs\\

           \If{$\mod({t}\times {f_{re}}) == 1$}{
                eVTOLs send the current physical information to the ATCo station\\
                The ATCo station sends the DT-deduced parameters to physical eVTOLs
           }

            $t = t + 1$  \\
    }
\end{algorithm}

The overall process of the proposed DT-based system is illustrated in Alg.~\ref{Ag5}. The process begins by using the centerline of the prescribed corridor $\mathcal{Q}$ as the initial eVTOL trajectory. This serves as the input for SIM communication optimization in $DT^S$, as the SIM optimization process requires a predefined trajectory.

In the first stage, $DT^S$ iteratively executes Alg.\ref{Ag2} and Alg.\ref{Ag3} to optimize the transmission power $\bm{P}$ and the phase shift $\bm{\Psi}$, respectively. 
These updated SIM parameters are then used by $DT^e$, which employs Alg.~\ref{Ag4} to determine optimized Composite Potential Field (CPF) hyperparameters ${k_{tar}, k_{sep}, k_{com}}$. These hyperparameters ensure safe and communication-efficient eVTOL aviation.

After completing the DT deduction iterations (Alg.\ref{Ag2}, Alg.\ref{Ag3}, and Alg.~\ref{Ag4}), the optimized SIM parameters $\bm{P}$ and $\bm{\Psi}$ are transmitted to the SIM antenna via wired fiber. Simultaneously, the CPF hyperparameters ${k_{tar}, k_{sep}, k_{com}}$ are sent to the corresponding eVTOL through SIM-based air-ground communication.

Despite these optimizations, practical challenges such as unexpected air obstacles (e.g., drones or birds) can introduce biases in the DT-deduced parameters during physical operations. To address this, the physical eVTOL platoon must continuously update its real-time dynamic data—such as velocity and position—to the ATCo station. This data is critical for revising any deduction bias in the DT-derived parameters.

Upon receiving the updated eVTOL data, $DT^e$ recalibrates its configurations to generate revised CPF hyperparameters, thereby compensating for any biases. These updated hyperparameters from $DT^e$ are subsequently used to refine the SIM optimization process within $DT^S$. Through this iterative process, $DT^S$ and $DT^e$ collaboratively produce updated SIM communication and eVTOL flight parameters to guide future operations of the physical system.

Finally, the frequency of DT synchronization, denoted as $f_{re}$, plays a crucial role in mitigating DT deduction biases. Higher synchronization frequencies can help maintain more accurate and reliable DT-derived parameters under dynamic conditions.


\section{Performance Evaluation}
To evaluate the effectiveness of the proposed scheme for eVTOL-based AAM, we conduct experiments on the SIM beamforming and CPF flight control methods. We first outline the simulation setup: the carrier frequency is set at $30$ GHz. The ATCo station, equipped with $M$ antennas, is positioned at the coordinate origin. 
Each SIM metasurface lies parallel to the $x$-$y$ plane, with its center aligned along the
$z$-axis. The SIM's thickness is defined as $Thick = 5\lambda$. For an $L$-layered SIM, the spacing between adjacent metasurfaces is $d_{Layer} = Thick / L$. 
Each metasurface consists of $K_x$ and $K_y$ meta-atoms along the $x$-axis and $y$-axis, respectively. 
Consequently, the total number of meta-atoms per layer becomes $K = K_x K_y$. For simplicity, we focus on the metasurface with a square structure where $K_x = K_y$. meaning each meta-atom's size is $d_x = d_y = \lambda / 2$. MATLAB is utilized to implement Alg.~\ref{Ag2} and Alg.~\ref{Ag3}, while PyTorch is adopted for Alg.~\ref{Ag4}. The primary simulation settings are summarized in Tab.~\ref{Simulation parameters}. 

\begin{table}[h]
\centering
\caption{Simulation parameters}
\begin{tabular}{|c|c|}
\hline
Number of time slots $N$ & 30 \\
\hline
Number of metasurface layers $L$ & 5 \\
\hline
Number of meta-atoms in each metasurface layer $K$ & 4 \\
\hline
Number of antennas $M$ & 3 \\
\hline
Wavelength $\lambda$ & 0.01 m \\
\hline
Total transmission power $P_{ATC}$ & 10 dBm \\
\hline
Reference channel gain $\rho_0$ & -30 dB \\
\hline
Noise power $\sigma_m^2$ & -90 dB \\
\hline
Path loss index $\alpha^{h}$ & 2.0 \\
\hline
Rician factor $\kappa^{h}$ & 10dB \\
\hline
\end{tabular}
\label{Simulation parameters}
\end{table}

\begin{figure} [h]
\centering
     \includegraphics[width=0.4\textwidth]{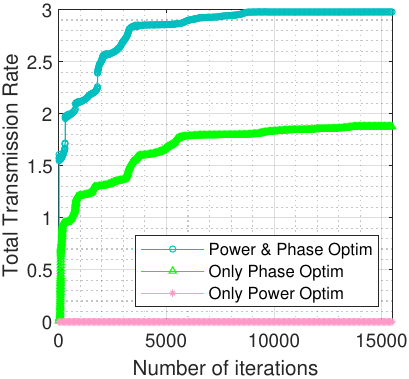} 
     \caption{Transmission rate comparison of different SIM communication optimizations} 
\label{SIM_optimization_Comparison}
\end{figure}

Fig.~\ref{SIM_optimization_Comparison} compares the transmission rates among three SIM optimization schemes for a given eVTOL trajectory. 
In the joint Power \& Phase Optimization scheme, the transmission rate is calculated by iteratively applying Alg.~\ref{Ag2} and Alg.~\ref{Ag3}.
In the Power-Only Optimization scheme, all phase shifts are initialized to $\theta_{k}^{l}[n] = 0$, and Alg.~\ref{Ag2} is then used to optimize the transmission rate.
Conversely, the Phase-Only Optimization scheme sets the transmission power of each antenna to $p_{m}[n] = 3.33$ dBm for every time slot and employs Alg.~\ref{Ag3} to optimize the transmission rate.

\begin{figure} [h]
\centering
     \includegraphics[width=0.48\textwidth]{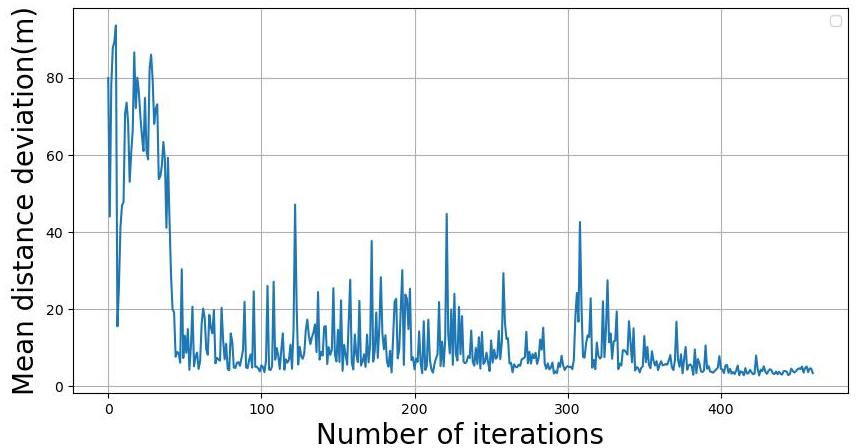} 
     \caption{Mean distance deviation of the eVTOL with the proposed DQN-based flight control} 
\label{DRLconvergence}
\end{figure}

As shown in Fig.~\ref{SIM_optimization_Comparison}, the joint Power \& Phase Optimization scheme achieves the best performance, with a transmission rate of approximately $3.0$ bps. 
The second-best performance is seen in the Phase-Only Optimization scheme, where the transmission rate converges to $1.9$ bps. 
On the other hand, the Power-Only Optimization scheme delivers the worst results, with rates approaching $0$ bps. 
This significantly highlights the importance of phase shifts in beam alignment. Relying solely on power optimization is insufficient for maintaining effective beam tracking for dynamic eVTOL trajectories.

\begin{figure} [h]
\centering
\includegraphics[width=0.4\textwidth]{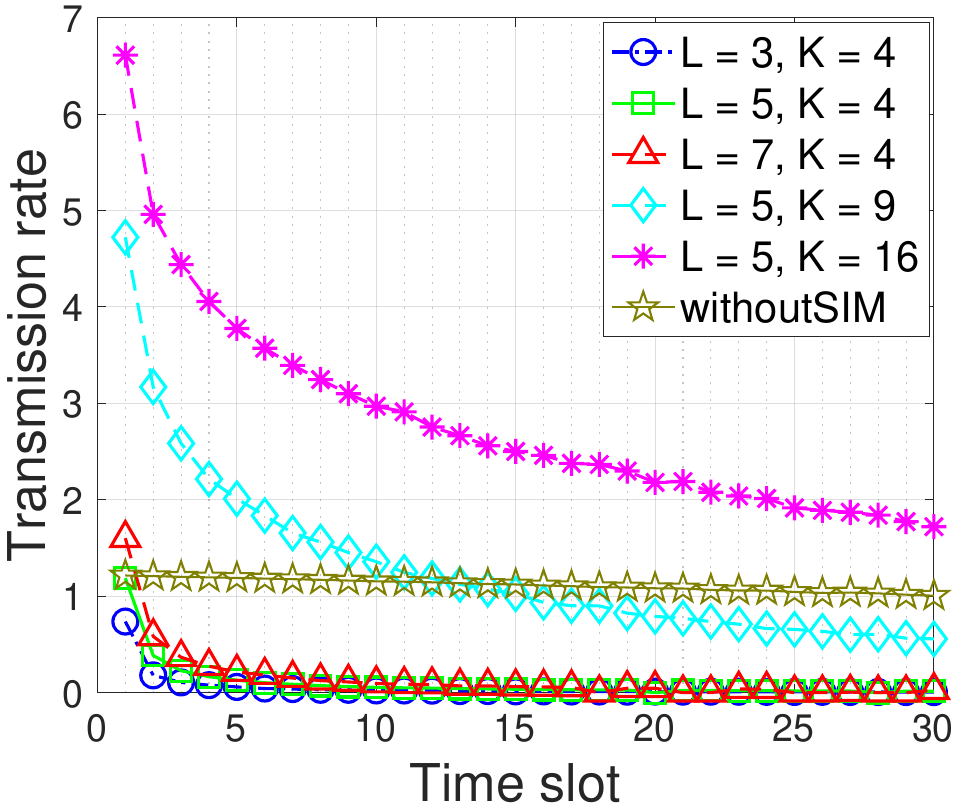} 
     \caption{Transmission rate of SIM v.s.  that  of without SIM} 
\label{SIM_unSIM_Comp}
\end{figure}

Fig.~\ref{DRLconvergence} illustrates the evolution of the proposed DQN-based CPF flight control algorithm over $500$ iterations. 
The $x$-axis represents the number of iterations, while the $y$-axis shows the mean distance deviation, defined as the average distance of each eVTOL from the corridor centerline.

Initially, the mean distance deviation decreases rapidly, reaching approximately $10$m by iteration 50. Beyond iteration 400, the deviation stabilizes at around $4$m, indicating that the eVTOLs have achieved a steady state and remain within the air corridor boundaries. 
This result demonstrates the convergence and performance efficiency of the proposed DQN-based CPF algorithm. With its rapid convergence and stable plateau, the CPF algorithm is well-suited for AAM applications.
Note that the converged $4$m deviation is achieved in the corridor without obstacles.
When there are obstacles in the corridor, the converged deviation will increase. 
It will be verified in the following simulation experiments.

\begin{figure} [h]
\centering
\includegraphics[width=0.44\textwidth]{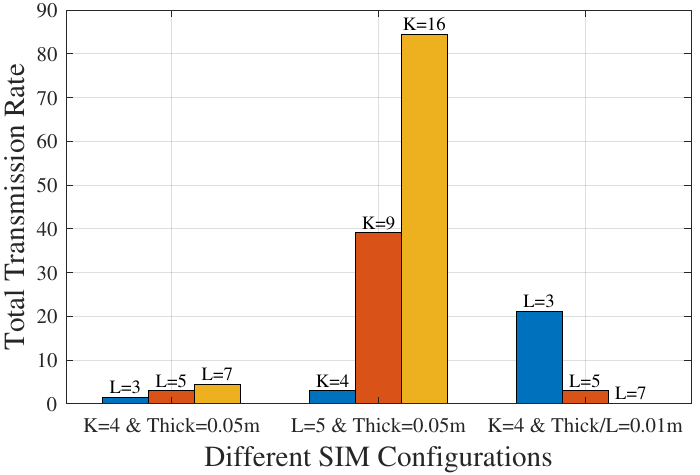} 
     \caption{Transmission rate v.s. different SIM configurations} 
\label{diff_L_K}
\end{figure}

Fig.~\ref{SIM_unSIM_Comp} compares the transmission rates over time for three eVTOLs, using SIM-based and MIMO-based ATCo stations. The SIM-based station includes five cases with varying configurations. For the MIMO-based ATCo station, the transmission capacity is given by:

\begin{equation}
\sum_{n=1}^N \sum_{m=1}^M \log \left(1+\frac{h_m[n]^2 p_m[n]}{\sum_{m^{\prime}=1, m^{\prime} \neq m}^M h_m[n]^2 p_{m^{\prime}}[n]+\sigma_m^2}\right),
\label{dsajfheffye}
\end{equation}

\noindent where the eVTOL trajectory is predetermined, and the transmission power is optimized using Alg.~\ref{Ag2}. For the SIM-based station, both transmission power and phase shifts are optimized iteratively using Alg.~\ref{Ag2} and Alg.~\ref{Ag3} with the same eVTOL trajectory.
As eVTOLs move farther from the ATCo station, the transmission rate decreases for both configurations. However, the SIM transmission rate declines faster than that of the MIMO transmission. This is due to the SIM phase shift amplifying the effect of distance on channel capacity. Among SIM configurations, the transmission rate of the metasurface setups ($L=5,K=9$ or $L=5,K=16$ where $L$ is the number of layers and $K$ is the total number of the meta-atoms per layer) outperforms that of the MIMO (without SIM). In contrast, the transmission rate of the metasurface setups ($L=3, K=4$, $L=5, K=4$, and $L=7, K=4$) is even worse than that of the station without SIM.

Moreover, Figure~\ref{diff_L_K} shows the variation of the total transmission rate with respect to the number of metasurface layers $L$, the number $K$ of meta-atoms per layer, and the total SIM thickness ($\textit{Thick}$).
For a fixed $K = 4$ and $\textit{Thick} = 0.05$m, the transmission rate increases with the number of metasurface layers $L$.
Similarly, for a fixed $L$ and $\textit{Thick}$, increasing the number of meta-atoms per layer can also improve the SIM transmission rate.
When the spacing between adjacent metasurface layers $\textit{Thick}/L$ is held constant (e.g., $\textit{Thick}/L = 0.01$m), increasing $L$ results in a thicker SIM and a corresponding decrease in transmission rate. 
This highlights a key limitation: SIM manufacturing constraints that impose a minimum gap between metasurface layers can reduce performance as $L$ increases.
For example, when $K = 4$ and $L = 3$, the transmission rate at $\textit{Thick}/L = 0.05$ is significantly lower ($1.5$ bps) than at $\textit{Thick}/L = 0.01$ m (about $21.2$ bps).
These results demonstrate that the transmission rate is inversely proportional to the distance between adjacent layers. Additionally, for a fixed $textit{Thick}$, increasing the total number of meta-atoms ($L \times K$) generally results in higher transmission rates.

For instance, the transmission rate for $K = 4$ and $L = 7$ ($K\times L=28$) is significantly lower than that for $K = 9$ and $L = 5$ ($K\times L=45$), yet higher than that for $K = 4$ and $L = 5$ ($K\times L=20$). 

\begin{figure}
\centering
\subfloat[DQN-based CPF trajectory]{\label{Conv}{\includegraphics[width=0.48\linewidth]{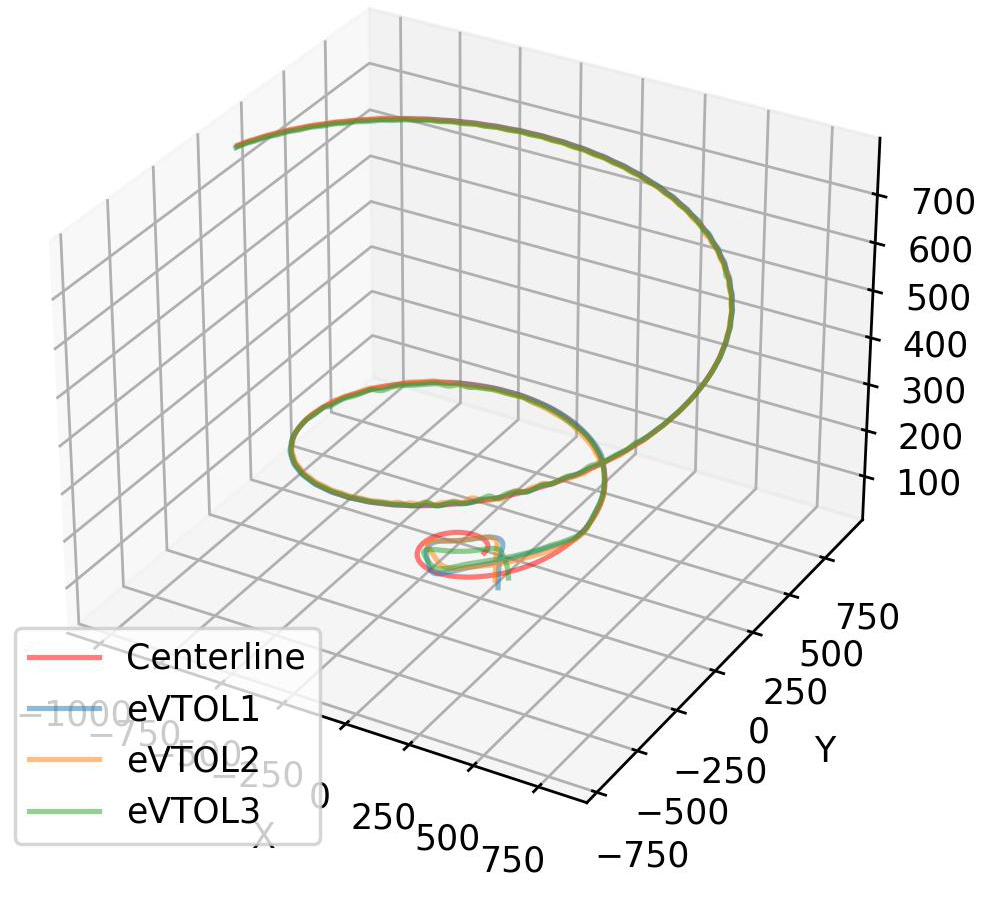}}}
\subfloat[CPF without DQN]{\label{UnConv}{\includegraphics[width=0.48\linewidth]{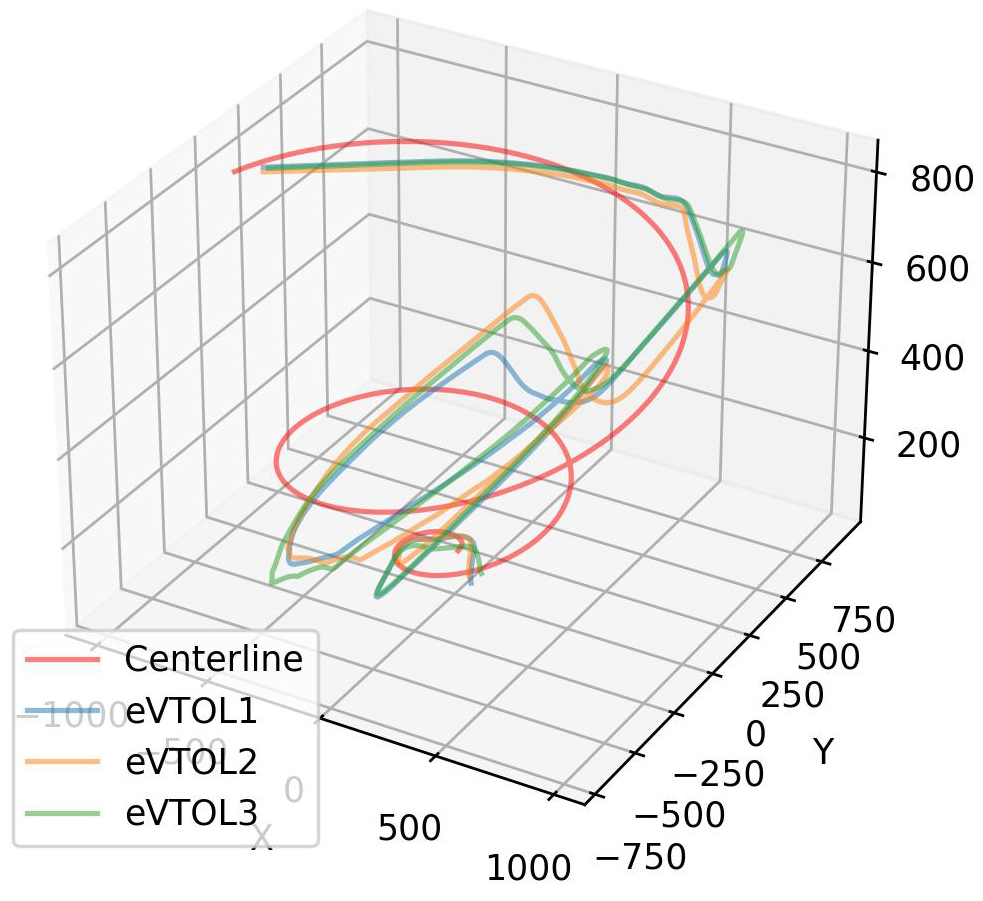}}}
\hfill 
\caption{DQN-based CPF method can maintain the eVTOL trajectory in the corridor}
\label{fggrgrgrgieit}
\end{figure}

Figure~\ref{fggrgrgrgieit} highlights the effectiveness of the DQN-based CPF method in maintaining eVTOL trajectories within the prescribed air corridor. The DQN-based trajectory shows fewer jitters compared to the original CPF method, providing smoother and safer flight paths. This improvement is essential for ensuring airspace safety and enhancing passenger experience.

\begin{figure}
\centering
\subfloat[Transmission rate comparison]{\label{TransmissionRateComparsion}{\includegraphics[width=0.45\linewidth]{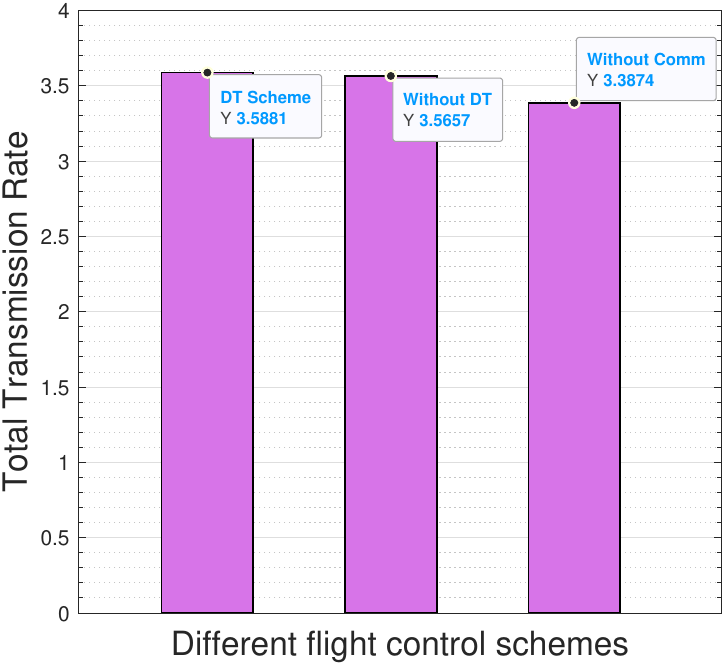}}}
\subfloat[DT-based eVTOL Trajectory]{\label{DTcomphy}{\includegraphics[width=0.5\linewidth]{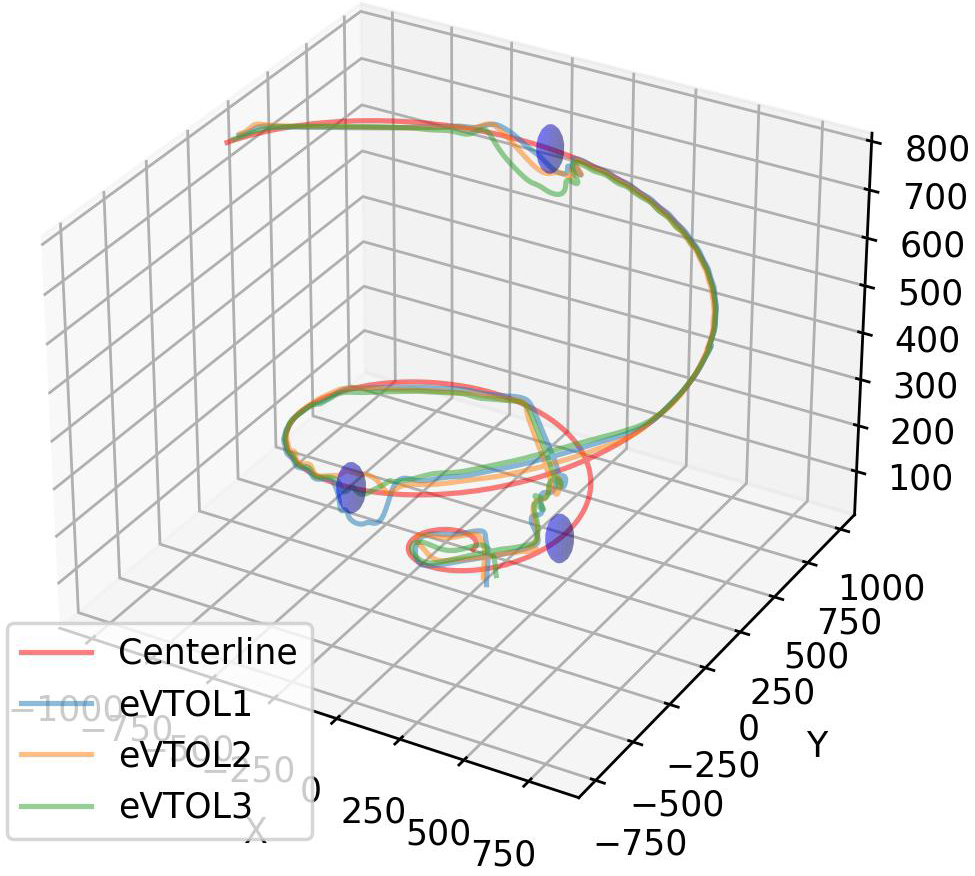}}}
\hfill 
\subfloat[Trajectory without DT-Sync]{\label{comphy}{\includegraphics[width=0.5\linewidth]{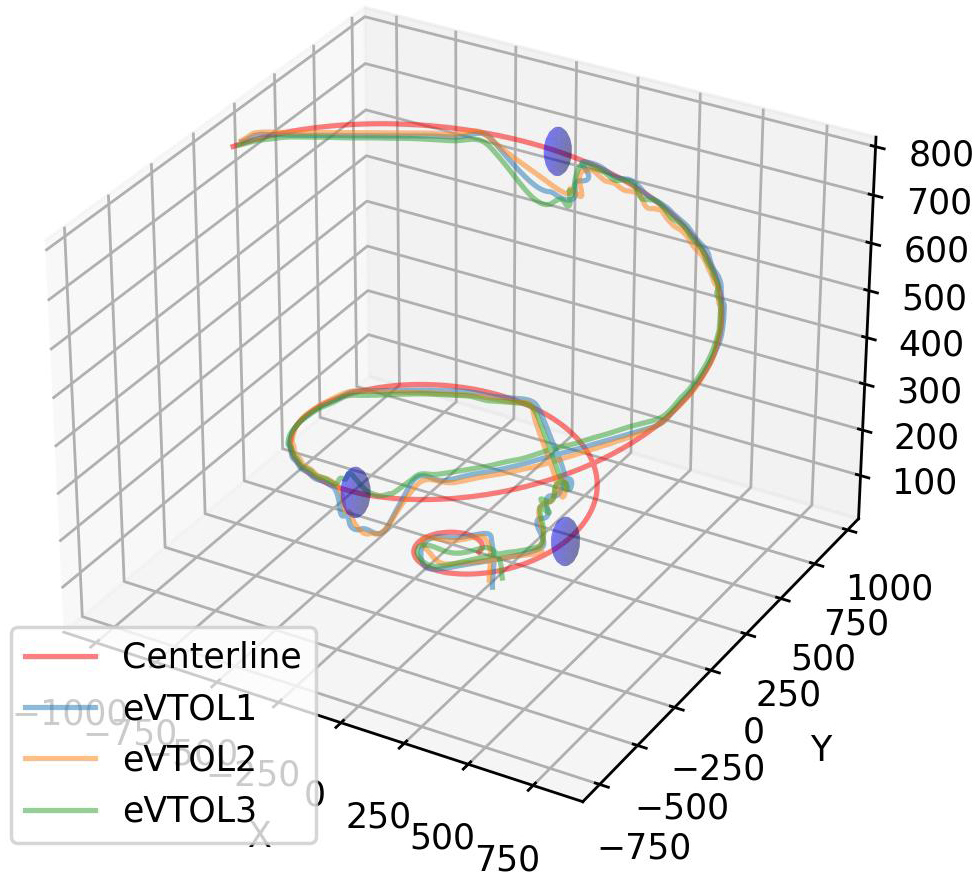}}} 
\subfloat[Trajectory without Comm]{\label{nocomphy}{\includegraphics[width=0.5\linewidth]{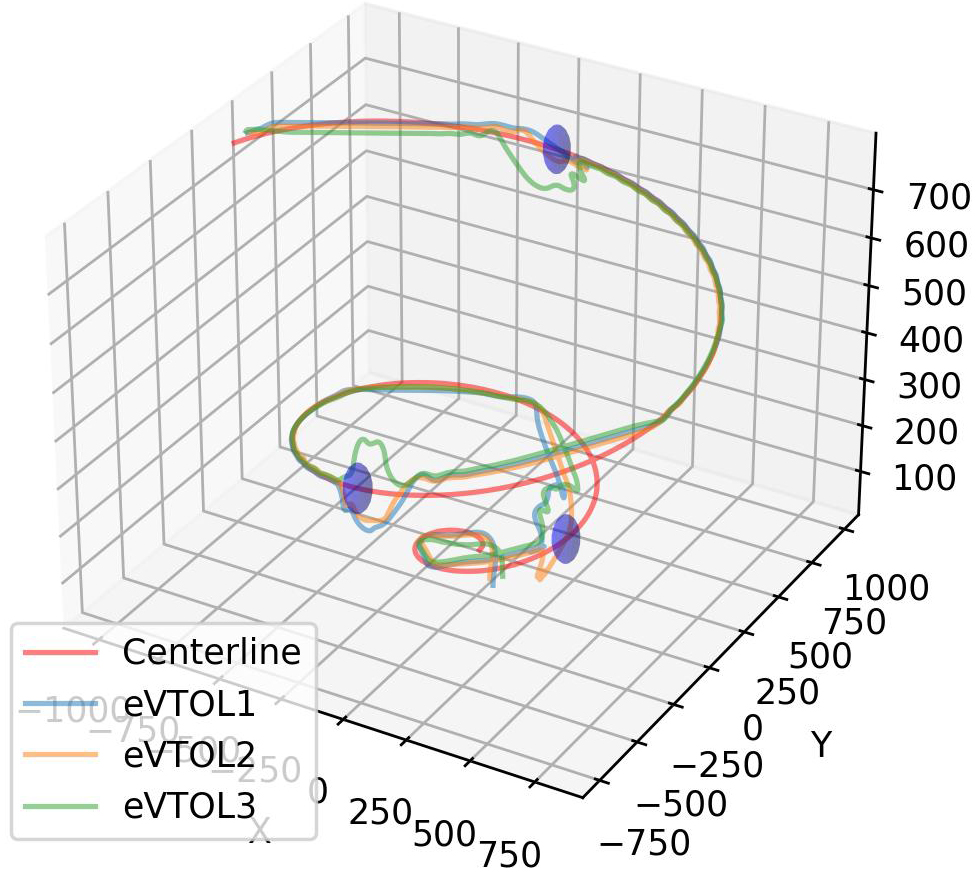}}}
\hfill 
\caption{Transmission rate with different flight control schemes}
\label{dsfguyihertugearuydfvguydsaz}
\end{figure}

Fig.~\ref{dsfguyihertugearuydfvguydsaz} compares the transmission rate and flight trajectories across three flight control schemes: the proposed DT scheme, the predetermined scheme (without DT synchronization), and the safe aviation-only scheme (without communication optimization). The simulation spans $30$ time slots, illustrating the flight process through the prescribed corridor.
In the proposed DT scheme, Alg.~\ref{Ag5} is executed with the DT synchronization procedure performed every 10 time slots. In contrast, the predetermined scheme without DT synchronization runs Alg.~\ref{Ag5} only once before the eVTOL platoon enters the corridor. 
This approach omits the DT synchronization step (lines 12–14 in Alg.~\ref{Ag5}), effectively generating a static, pre-optimal flight trajectory. 
Lastly, the scheme without communication optimization eliminates the communication component from Eq.~\ref{kbuycxvguiyesrjhsdfg}. This reduces the whole algorithm to a DQN procedure (Alg.~\ref{Ag4}) where the reward function is modified as: $r_i = \alpha_1 v_{tar}(1 + \alpha_2 d_{i,tar}) + \sum_{j \neq i, \, d_{i,j} < d_{sep}} \left(\frac{\beta v_{sep}}{d_{i,j}}\right)$.

Fig.~\ref{TransmissionRateComparsion} shows that the proposed DT scheme achieves the highest transmission rate among the three approaches. Conversely, the scheme without communication optimization performs the worst due to its lack of communication-oriented optimization. However, the overall performance differences between the three schemes remain relatively small. This is primarily because the air corridor's constraints limit the flexibility of flight trajectories, restricting additional gains in communication through mobility.
Fig.~\ref{DTcomphy}, \ref{comphy}, and \ref{nocomphy} depict the eVTOL flight trajectories under the DT scheme, the scheme without DT synchronization, and the scheme without communication optimization, respectively. In these figures, the blue entity represents a spherical obstacle with a radius of $50$ m. The DT scheme demonstrates a noticeably smoother trajectory compared to the other two schemes. 
This smoothness arises from the frequent DT synchronizations, which allow timely updates to the eVTOL flight parameters (CPF parameters), mitigating trajectory deviations caused by obstacles. In contrast, the other two schemes lack this DT synchronization mechanism, leaving the flight parameters static throughout the simulation, which cannot revise the trajectory deviation caused by obstacles.

\begin{figure}
\centering
\subfloat[Transmission rate comparison]{\label{5o_TransmissionRateComparsion}{\includegraphics[width=0.49\linewidth]{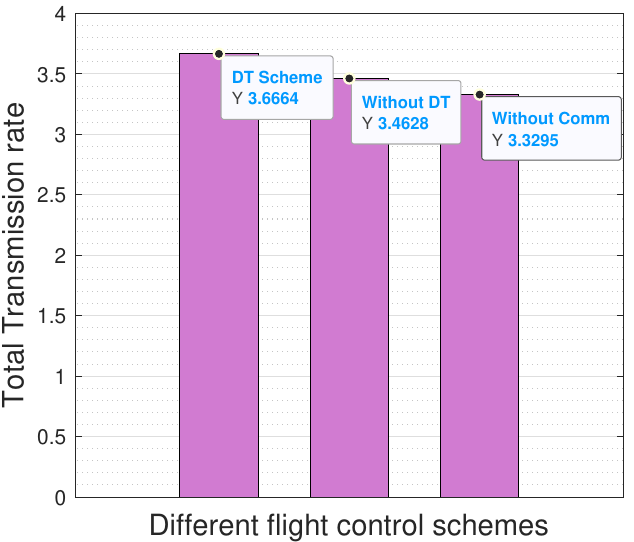}}}
\subfloat[DT-based eVTOL Trajectory]{\label{5o_DTcomphy}{\includegraphics[width=0.5\linewidth]{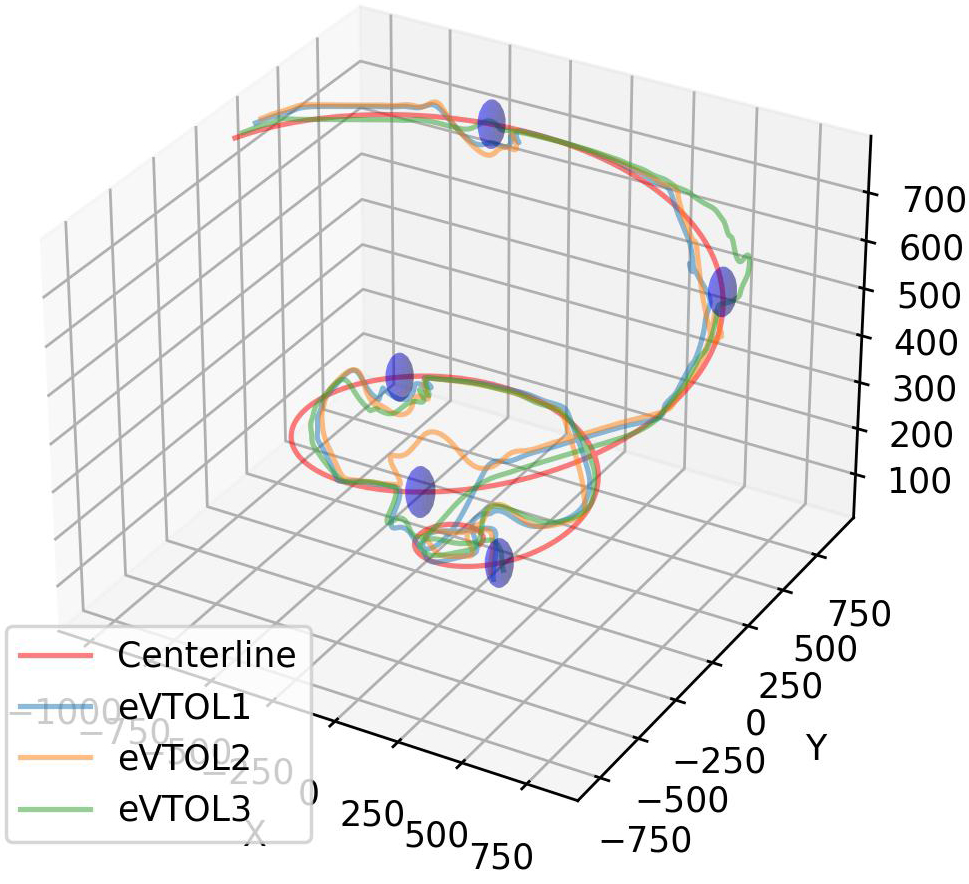}}}
\hfill 
\subfloat[Trajectory without DT-Sync]{\label{5o_comphy}{\includegraphics[width=0.5\linewidth]{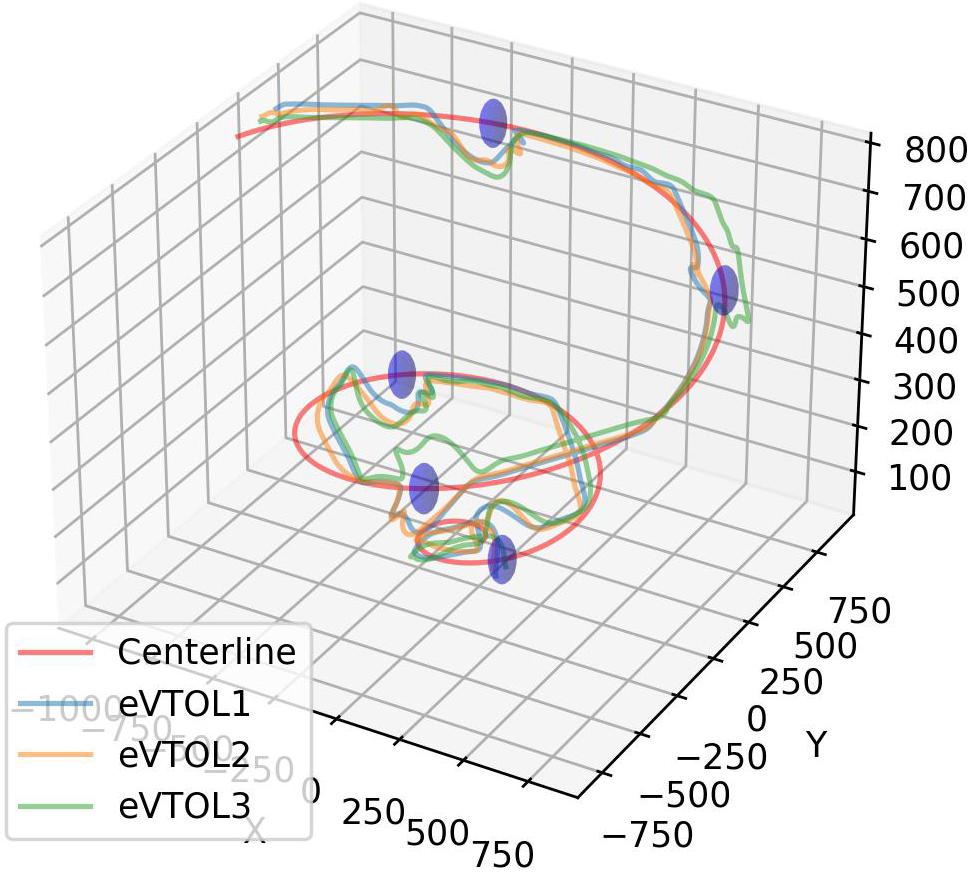}}} 
\subfloat[Trajectory without Comm]{\label{5o_nocomphy}{\includegraphics[width=0.5\linewidth]{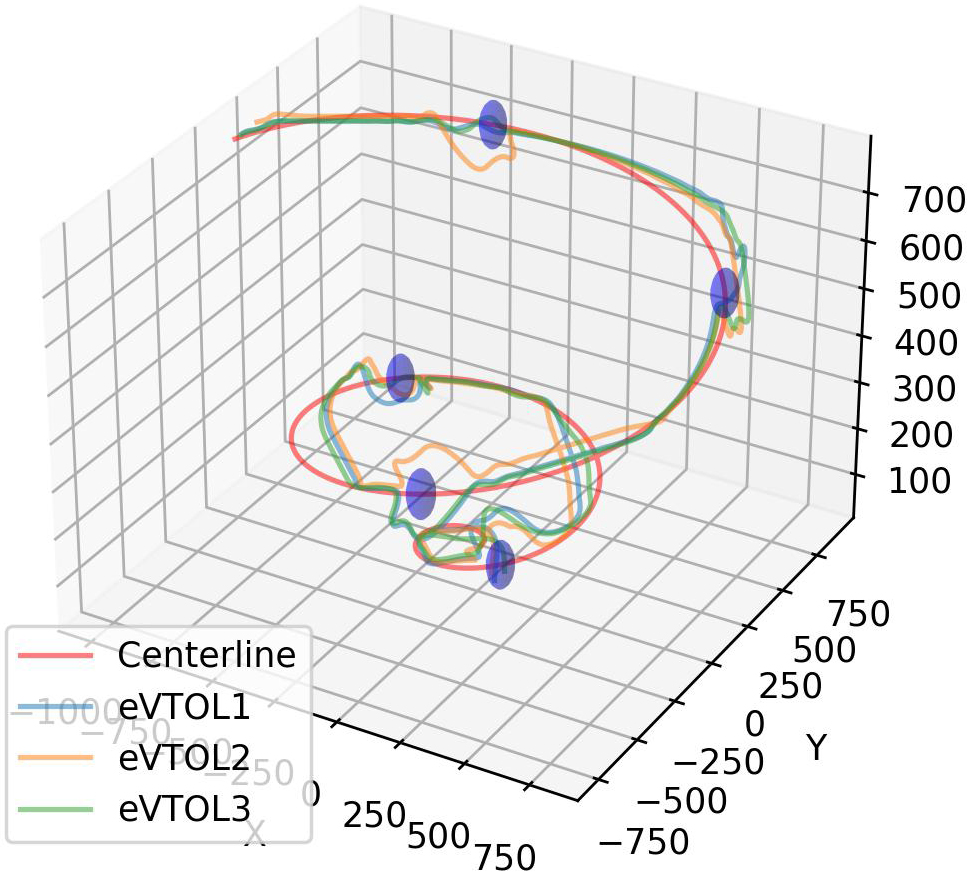}}}
\hfill 
\caption{Different flight control schemes with 5 obstacles}
\label{5o_dsfguyihertugearuydfvguydsaz}
\end{figure}

Fig.~\ref{5o_dsfguyihertugearuydfvguydsaz} examines the performance of the flight control schemes in the air corridor containing five obstacles. All schemes successfully navigated in the corridor without colliding with obstacles, showcasing the robustness of the proposed DQN algorithm. Moreover, the proposed DT scheme achieves the highest transmission rate, outperforming the other approaches.
Compared to the CPF flight control benchmark, it improves the transmission rate by 8.3\%. 
When comparing scenarios with three obstacles versus five, the performance gap between the DT-based scheme and the predetermined scheme widens. This is because DT synchronization can revise deviations caused by obstacles. 
As the number of obstacles increases, the frequency of trajectory deviations and jitters also grows, amplifying the benefits of DT synchronization. Conversely, flight control schemes without DT synchronization suffer from reduced performance due to their inability to adapt to dynamic environmental changes.

\begin{figure}
\centering
\subfloat[Transmission rate]{\label{rate_DTsync}{\includegraphics[width=0.45\linewidth]{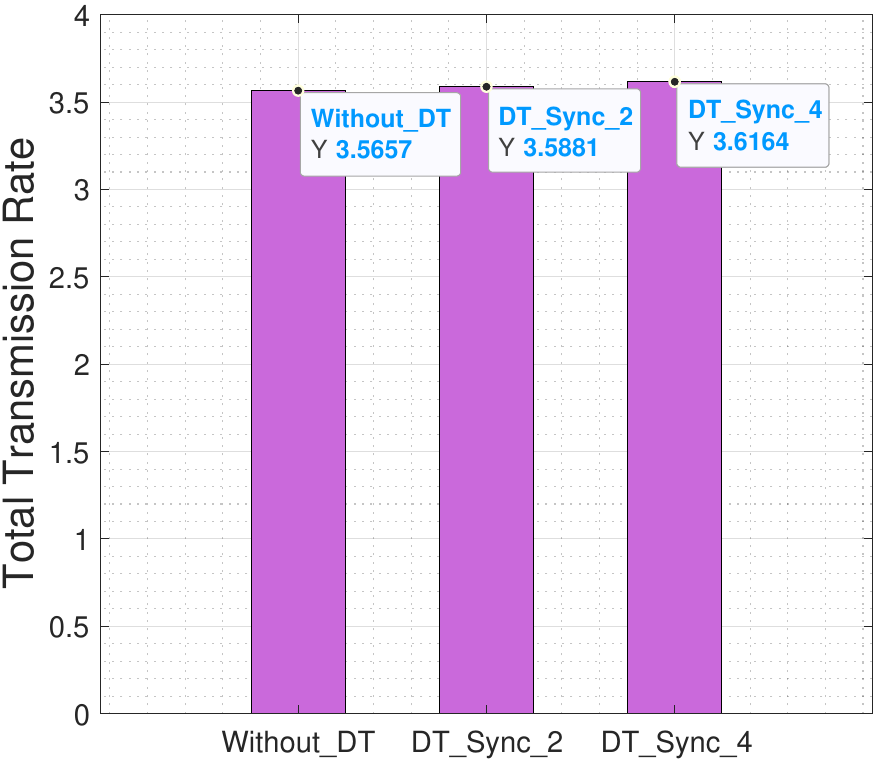}}}
\subfloat[Mean distance deviation]{\label{dis_DTsync}{\includegraphics[width=0.5\linewidth]{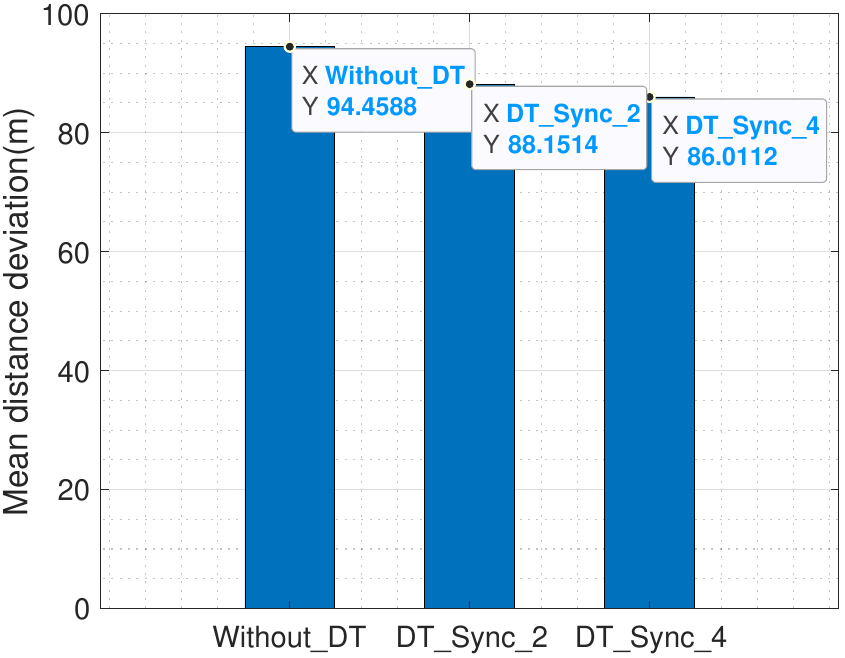}}}
\hfill
\caption{Impact of DT synchronization in a corridor with three obstacles}
\label{sbhursbdoaiufgeraufv}
\end{figure}

Fig.~\ref{sbhursbdoaiufgeraufv} highlights the impact of DT synchronization frequency on transmission rates and mean distance deviations in a corridor with three obstacles.
Note that the DT synchronization procedure is realized as Line 10 to 11 of Alg.~\ref{Ag5}. 
Fig.~\ref{rate_DTsync} shows that performing DT synchronization four times results in slightly better transmission rates than synchronizing twice or not at all. 
It reduces flight distance deviation from the prescribed corridor by 10\% compared to the predetermined optimization (without DT).
However, this improvement is modest because the optimization freedom within a corridor with three obstacles is inherently limited. Even a pre-scheduling scheme (without DT) can handle this scenario adequately.
In contrast, Fig.~\ref{dis_DTsync} demonstrates a significant reduction in mean distance deviation as the frequency of DT synchronization increases. Frequent synchronization allows eVTOLs to promptly correct flight deviations, ensuring that their trajectories remain closer to the corridor centerline. This finding underscores the importance of DT synchronization for precise flight control, particularly in complex or obstacle-rich environments.

\section{Conclusion}
This paper presents a Digital Twin (DT)-based flight control scheme for eVTOLs, aimed at ensuring both safe and communication-efficient aviation within prescribed air corridors. 
The proposed scheme comprises three sub-optimization algorithms, each addressing a specific objective: transmission power optimization, SIM phase shift optimization, and flight control optimization.
On the DT side, the transmission power and phase shift optimization algorithms work iteratively within the SIM DT. 
The results from the SIM DT are then fed into the eVTOL DT. 
Using these pre-optimized SIM parameters, the eVTOL DT applies a DQN-based algorithm to refine the CPF hyperparameters for safe aviation. 
These hyperparameters play a crucial role in generating subsequent trajectory for the next communication optimization of the SIM DT.
Therefore, the DT process involves multiple iterations between the SIM DT and the eVTOL DT. 
Through this iterative exchange, the final DT outcomes—including optimized transmission power, phase shift values, and potential field hyperparameters—are generated. 
These parameters are then transmitted to the physical SIM antenna and eVTOLs. On the other side, the physical entities will regularly update their operation parameters to the DT side, for seamless synchronization between the digital and physical systems.
Extensive experimental results validate the effectiveness and superiority of the proposed scheme compared to existing benchmarks. The approach not only improves communication efficiency but also significantly enhances flight safety along the prescribed air corridor.
In future work, research will focus on expanding the framework to support large-scale eVTOL swarm operations across multiple prescribed air corridors.

\appendices
\footnotesize
\bibliography{biblio}

\end{document}